\documentclass[10pt,preprint,5p,twocolumn,times]{elsarticle}

\usepackage{graphicx}
\usepackage{hyperref}
\usepackage{rotating}
\usepackage{amsmath}
\usepackage{amssymb}
\usepackage{subfig}
\usepackage{color}
\usepackage{algorithm}
\usepackage{algpseudocode}

\renewcommand{\vec}[1]{{\bf #1}}

\graphicspath{
  {./}
}
\begin{document}
\journal{Computer Physics Communications}
\begin{frontmatter}
\bibliographystyle{model1a-num-names}
\biboptions{square,numbers,comma,sort&compress}

\title{Efficient Parallel Linear Scaling Method to get the Response Density Matrix in All-Electron Real-Space Density-Functional Perturbation Theory}

\author[CAS]{Honghui Shang}
\author[XU]{Wanzhen Liang}
%\ead{liangwz@xmu.edu.cn}
\author[CAS]{Yunquan Zhang}
\author[USTC]{Jinlong Yang}
%\ead{jlyang@ustc.edu.cn}

\address[CAS]{State Key Laboratory of Computer Architecture, Institute of Computing Technology, Chinese Academy of Sciences, Beijing, 100190, China}
\address[XU]{State Key Laboratory of Physical Chemistry of Solid Surfaces, Collaborative Innovation Center of Chemistry for Energy Materials, Fujian Provincial Key Laboratory of Theoretical and Computational Chemistry, and Department of Chemistry, College of Chemistry and Chemical Engineering, Xiamen University, Xiamen, Fujian 361005, People’s Republic of China}
\address[USTC]{Hefei National Laboratory for Physical Sciences at Microscale, Department of Chemical Physics, and Synergetic Innovation Center of Quantum Information and Quantum Physics, University of Science and Technology of China, Hefei, Anhui 230026, China }

%%%%%%%%%%%%%%%%%%%%%%%%%%%%%%%%%%%%%%%%%%%%%%%%%%%%%%%%%%%%%%%%%%
%                            Abstract                            %
%%%%%%%%%%%%%%%%%%%%%%%%%%%%%%%%%%%%%%%%%%%%%%%%%%%%%%%%%%%%%%%%%%
\begin{abstract}
The real-space density-functional perturbation theory (DFPT) for the computations of the response properties with respect to the atomic displacement and homogeneous electric field perturbation has been recently developed and implemented into the all-electron, numeric atom-centered orbitals
electronic structure package FHI-aims.
It is found that the bottleneck for large scale applications is the computation of the response density matrix, which scales as $O(N^3)$.
Here for the response properties with respect to the 
homogeneous electric field, we present an efficient parallel linear scaling algorithm for the response density matrix calculation.
Our scheme is based on the second-order trace-correcting purification and the parallel sparse matrix-matrix multiplication algorithms. The new scheme reduces the formal scaling from $O(N^3)$ to $O(N)$, and shows good parallel scalability over tens of thousands of cores. As demonstrated by extensive validation, we achieve a rapid computation of
accurate polarizabilities using DFPT. Finally, the computational efficiency of this scheme has been illustrated by making the scaling tests and scalability tests on massively parallel computer systems. 
\end{abstract}
\end{frontmatter}

\section{Introduction}
%------something about DFPT---------------
Density-functional theory (DFT) \cite{Hohenberg1964,Kohn1965} 
 applied in chemistry, physics, and material science is the ground-state theory through which one can calculate the total energy and its first order derivatives~(e.g. dipole moment and force).  The response properties (e.g., polarizability, vibrational frequencies or phonon dispersions) related to the second and higher order derivatives of the total energy can be obtained within the same framework by means of density-functional perturbation theory (DFPT) \cite{Gonze1997-1,Gonze1997-2,Baroni-2001} or the so-called coupled perturbed self-consistent field (CPSCF) method\cite{Gerratt-1967,Pople-1979,Dykstra-1984,Frisch-1990,
Ochsenfeld-1997, Liang-2005} in the quantum chemistry community.

Recently, we have developed and implemented a real-space formalism for DFPT~\cite{Shang2017} in the all-electron, full-potential, numerical atomic orbitals based Fritz Haber Institute {\it ab initio} molecular simulations (FHI-aims) package, which allows us to take advantage of the inherent locality of the basis set to achieve a numerically favorable scaling. Such real-space DFPT has been applied in lattice dynamics calculations\cite{Shang2017} and in computations of the polarizabilities, dielectric constants, harmonic as well as anharmonic Raman spectra~\cite{Shang2018}, in which good 
computational accuracy, computational efficiency, and parallel scalability have been demonstrated.

%-------the tradition ON3--------------- 
In our previous scaling test~\cite{Shang2017,Shang2018}, it was found that, for large systems with more than 1,000 atoms, the computational cost for updating the response density matrix becomes dominant. It is because the dense matrix multiplication operations in this step scale as~$O(N^{3})$, and thereby, presents a serious bottleneck to deal with large systems. It is desirable to make the computational time scale linearly,~i.e.~$O(N)$, with the size of the system~\cite{Bowler2011}.
%-------how to reduce the ON3 to ON1-----
In order to achieve this goal, the Kohn’s nearsightedness principle~\cite{Kohn1996} need to 
be adopted. It says that, for a quantum mechanical system within an external potential, its local properties do not ``see'' a change of the external potential if this change is limited to a distant region. This fundamental principle is behind almost all linear scaling algorithms\cite{Soler2002,Bowler2010,Shang2010a,CONQUEST,Weber2006,Wu2009,Bowler2010}, which leads to the sparsity of the density matrix $-$ a key to achieve the linear scaling. 
%In practice, the sparse matrix is achieved by dropping matrix elements less than a threshold or by using strictly truncated atomic basis sets. 
Using sparse zero-order density and Hamiltonian matrices, in 2002, Niklasson suggested a trace-correcting~(TC2) approach to replace the traditional  diagonalization step through a density matrix purification method \cite{Niklasson2002PRB}.
Later in 2004, Niklasson and Challacombe proposed the density matrix perturbation theory~(DMPT)~\cite{Niklasson2004}  to extend the TC2 approach in calculating response density matrices. In contrast to the traditional density functional perturbation theory, where the first-order density matrix is calculated from dense eigenstate coefficients matrices, the DMPT approach only adopts sparse first-order Hamiltonian and density matrices, leading to the linear scaling of the calculations. Such DMPT approach can be further combined with the CPSCF cycles, and in this way, the response properties can be calculated self-consistently. This combination is called the TC2-CPSCF method throughout this paper.

The advantage of linear scaling in calculations can be significantly enhanced if the computations are performed in a massively parallel way. Currently, the parallelization of the TC2 method has been achieved in a few different ways, like using in-node parallelism via multithreading scheme within one node~\cite{Mniszewski2015,Cawkwell2014}, using MPI parallelization based on the distributed block compressed sparse row~(DBCSR) library~\cite{Lazzaro2017} in CP2K~\cite{Hutter2014}, or using the hybrid MPI+OpenMP parallelization scheme~\cite{Azad2016} for the coordinate~(COO) data format\cite{Dawson2018}. For the TC2-CPSCF method, however, there is only serial implementations~\cite{Weber2004,Xiang2006} and no parallel implementation and scalability performance testing has been done yet.

%-------what we do in this paper------
In this work, we have implemented the sparse matrices-based TC2-CPSCF method in the FHI-aims package. We have also parallelized the code by using the MPI level distributed memory parallelization algorithm. The linear scaling and good parallel scalability have been achieved for the response density matrix calculation within DFPT. The linear scaling with system sizes up to several thousands of atoms and the scalability on tens of thousands of cores are demonstrated using various realistic systems.

%------outline-------------
The remainder of this paper is organized as follows. The 
fundamental theoretical framework is presented in Sec.~\ref{sec:method}.
In Sec.~\ref{sec:result}, the implementation is validated by comparing the calculated analytical polarizabilities with results obtained from the traditional $O(N^{3})$ approach. We also discuss the convergence behavior of the implementation, the scaling of the computational cost with the system sizes, and the parallel performance on a large number of cores. In Sec.~\ref{sec:conclusions}, we summarize our main achievement and highlight the relevance of this work to the parallel implementation of other methods.

\section{Method}
\label{sec:method}
In this section, the basic equations are introduced. Here we use 
a spin-unpolarized notation for the sake of simplicity, a formal generalization to spin-polarized notation is straightforward. 
Moreover, in this work, we focus on the equations for finite systems, the generalization to extended periodic solid case can be found in our previous work\cite{Shang2017,Shang2018}. 

In Kohn-Sham density-functional theory, the total energy is uniquely determined by the electron density $n(\mathbf{r})$  
\begin{eqnarray} 
  E_{tot}     & = & \underbrace{-\dfrac{1}{2}\sum_{i}<\psi_i|\nabla^2|\psi_i> }_{T_{s}[n]}     -  \underbrace{ \int {n(\mathbf{r})  \sum_{I}\dfrac{Z_{I}}{|\mathbf{r}-\mathbf{R}_{I}|}  d\mathbf{r}} }_{ E_{ext}[n]}  \nonumber   \\  
  & + & \underbrace{ \dfrac{1}{2}\int \int {\dfrac{n(\mathbf{r}) n(\mathbf{r'}) }{|\mathbf{r}-\mathbf{r'} |}  d\mathbf{r}  d\mathbf{r'}} }_{E_{H}[n]}  +   \underbrace{ \dfrac{1}{2}\sum_{I}\sum_{J}{\dfrac{Z_{I} Z_{J}}{|\mathbf{R}_{I}-\mathbf{R}_{J} |} }    }_{E_{ion-ion}} \nonumber   \\
  & + & E_{xc}[n] \;,
\label{eq:KSTOT}
\end{eqnarray}
in which $\psi_i$ is the Kohn-Sham eigenstate, $T_{s}$ is the kinetic energy of non-interacting electrons, $E_{ext}$ the electron-nuclear, $E_H$ the Hartree, $E_{xc}$ the exchange-correlation, and $E_{ion-ion}$ the ion-ion repulsion energy. All energies are functionals of the electron density. The electron density is written with the eigenfunction,
\begin{equation}
n(\mathbf{r})=\sum_i f_i |\psi_i(\mathbf{r})|^2|,
\end{equation}
in which $f_i$ denotes the occupation number of eigenstate~$\psi_i$.  The ground state electron density~$n(\vec{r})$ 
is obtained by variationally minimizing Eq.~(\ref{eq:KSTOT})
\begin{equation}
\dfrac{\delta}{\delta n}\left[E_{tot}- \mu\left(\int \! \! n(\vec{r}) \: d\vec{r} - N_e\right) \right] 
 =  0 \;,
\label{eq:ks-variational}
\end{equation}
in which $\mu=\delta E_{KS}/\delta n$ is the chemical potential. From above equation we get the Kohn-Sham~(KS) single particle equations
\begin{equation}
\hat{h}_{KS}\psi_i = \left[ \hat{t}_s + \hat{v}_{ext}(r)+\hat{v}_{H}+\hat{v}_{xc}\right] \psi_i = \epsilon_{i} \psi_i \;,
\label{eq:ks-equation}
\end{equation}
for the Kohn-Sham Hamiltonian~$\hat{h}_{KS}$,  in which, $\hat{t}_s$ is the single particle kinetic operator, $\hat{v}_{ext}$ the
(external) electron-nuclear potential, $\hat{v}_{H}$ the Hartree potential, and $\hat{v}_{xc}$ the exchange-correlation potential. The Kohn-Sham single particle states~$\psi_i$ and their eigenenergies~$\epsilon_{i}$ can be calculated by solving Eq.~(\ref{eq:ks-equation}).
In practical numerical implementations, the Kohn-Sham states~$\psi_i$ are expanded with the finite basis set~$\chi_\mu(\vec{r})$
\begin{equation}
\psi_i(\mathbf{r})=\sum_{\mu}C_{\mu i} \: \chi_{\mu}(\mathbf{r})\;,
\label{eq:expansion}
\end{equation}
using the expansion coefficients $C_{\mu i}$. Here the numeric atom-centered orbitals~(NAOs)\cite{Blum2009,Delley1991,Delley-partition} are adopted as the basis set $\chi_\mu(\vec{r})$. Denoting $H_{\mu\nu}=\int{\chi_{\mu}(\mathbf{r}) \hat{h}_{KS}  \chi_{\nu}(\mathbf{r}) d\mathbf{r}}$ for the Hamiltonian matrix and $S_{\mu\nu}=\int{\chi_{\mu}(\mathbf{r}) \chi_{\nu}(\mathbf{r}) d\mathbf{r}}$ 
for the overlap matrix, we can rewrite Eq.~(\ref{eq:ks-equation}) as 
\begin{equation}
\sum_{\nu} H_{\mu\nu} C_{\nu i} = \epsilon_{i}  \sum_{\nu} S_{\mu\nu} C_{\nu i}\;.
\label{eq:KS-element}
\end{equation}
And we can write it in the more convenient matrix form for the zero order Kohn-Sham equation: 
\begin{equation}
H^{(0)}C^{(0)}=  S^{(0)} C^{(0)}E^{(0)} \;,
\label{eq:KS-matrix}
\end{equation}
whereby $E^{(0)}$ denotes the diagonal matrices containing the eigenvalues~$\epsilon_i$.

%----------first-order------------
If an external electric field~$\vec{E}=\left(e_x,e_y,e_z\right)$ with  strengths~$e_\gamma$  is applied to an isolated system, the KS Hamiltonian gains an additional term~$\hat{h}_{E}= - \vec{r} \cdot \vec{E}$, which contributes
\begin{equation}
E_{E}[n] = - \sum_\gamma \int e_\gamma r_\gamma \, n(\vec{r}) \, d\vec{r} 
\end{equation}
to the total energy functional in Eq.~(\ref{eq:KSTOT}). 
A perturbative Taylor-expansion of the total energy in the zero-field limit gives
\begin{equation}
E_{tot}(\vec{E}) \approx E^0_{tot} - \sum_{\gamma}\mu_{\gamma}e_{\gamma} - \frac{1}{2}\sum_{\gamma,\delta}\alpha_{\gamma\delta}e_{\gamma}e_\delta + \cdots\;, 
\end{equation}
where $\delta,\gamma$ are Cartesian directions. For isolated systems, the coefficient in the linear term is
\begin{equation}
\mu_{\gamma}= \int{n_0(\vec{r}) r_\gamma d\vec{r} },
\label{eq:dipole}
\end{equation}
which corresponds to the $\gamma$-component of the dipole moment. The coefficient in the second-order term is the polarizability

\begin{equation}
\alpha_{\gamma\delta} 
 =    \frac{\partial \mu_{\gamma}}{\partial e_{\delta}} 
 =  \int{ r_\gamma  .\frac{\partial n_0(\vec{r})}{\partial e_{\delta} }d \vec{r}},
\label{eq:polar}
\end{equation}
which need to be calculated with the response of the 
ground-state density with respect to the field strength according to the $2n+1$ rule~\cite{Gonze-1989}. We use the notation $M^{(1)}$ for the first order response quantities with respect to the  homogeneous external electrical field. 
\begin{equation}
M^{(1)}=\frac{d {M^{(0)}}}{d {e_{\gamma}}} \; .
\end{equation}
Then the first order response of Eq.~(\ref{eq:KS-element}) is written as
\begin{equation}
 \sum_{\nu}(H_{\mu\nu}^{(0)}-\epsilon_i^{(0)} S_{ \mu\nu}^{(0)})C_{\nu i}^{(1)}    
= -\sum_{\nu}\left( H_{\mu\nu}^{(1)} - \epsilon_i^{(1)}S_{\mu\nu}^{(0)}\right) C_{\nu i}^{(0)} \;.
\label{eq:Sh-element}
\end{equation}
It should be noted that, for the homogeneous external electrical field perturbation discussed in this work, the first order overlap matrix $S^{(1)}$ is zero since the overlap
matrix does not change with respect to the electrical field  perturbation.
And we can also have its matrix form :
\begin{equation}
H^{(0)}C^{(1)}-S^{(0)}C^{(1)}E^{(0)} =-  H^{(1)}C^{(0)} + S^{(0)}C^{(0)}E^{(1)}  \;, 
\label{eq:Sh-matrix}
\end{equation}
whereby $E^{(0)}$ and $E^{(1)}$ denote the diagonal matrices containing the eigenvalues~$\epsilon_i$ and their responses respectively. 
The Eq.~(\ref{eq:Sh-element}) and Eq.~(\ref{eq:Sh-matrix}) are called  Sternheimer equation~\cite{Sternheimer1954}, which is the key to get the response density matrix per CPSCF cycle in the density-functional perturbation theory\cite{Gonze1997-1,Gonze1997-2,Baroni-2001,Shang2017,Shang2018}.

\subsection{The $O(N^3)$ method to get the first order density matrix}
\label{subsec:ON3}
The traditional $O(N^3)$ way~\cite{Gerratt-1967,Pople-1979,Dykstra-1984} to get the first order density matrix
 in each CPSCF cycle includes two steps. Firstly
the Sternheimer equation~(Eq.~\ref{eq:Sh-matrix}) is solved to get
the first order coefficients ~$C^{(1)}$; Secondly, the response~(first
order) density matrix is constructed with the first order coefficients ~$C^{(1)}$ and the occupation number ($f_i$) of eigenstate
\begin{equation}
P_{\mu,\nu}^{(1)}=\sum_{i}{ f_i
\left(C_{\mu,i}^{(1)} C_{\nu,i}^{(0)} + C_{\mu,i}^{(0)} C_{\nu,i}^{(1)}    \right)  }  \;.
\label{eq:DM1}
\end{equation}
In the first step to solve the Sternheimer equation, the first order coefficients~$C^{(1)}$ are expanded in terms of the zero order expansion coefficients~$C^{(0)}$ using
\begin{equation}
C^{(1)}=C^{(0)}U^{(1)} \quad \text{i.e.}  ~ C^{(1)}_{\mu i}= \sum_{p}C^{(0)}_{\mu p} U^{(1)}_{pi} \;, 
\end{equation}
Then by multiplying Eq.~(\ref{eq:Sh-matrix}) with the Hermitian conjugate~$C^{(0)\dagger}$, and using the orthonormality relation,
\begin{equation}
C^{(0)\dagger}S^{(0)}C^{(0)}=1 \;,
\label{eq:orthonormal}
\end{equation}
we get
\begin{align}
\label{eq:CPSCF-U}
E^{(0)}U^{(1)}-U^{(1)}E^{(0)} & \\ = -C^{(0)\dagger}H^{(1)}C^{(0)}+E^{(1)} \;. \nonumber
\end{align}
Due to the diagonal character of~$E^{(0)}$ and~$E^{(1)}$,
this matrix equation contains the response of the eigenvalues on
its diagonal elements
\begin{equation}
\epsilon_p^{(1)}= \left[ C^{(0)\dagger}H^{(1)}C^{(0)} \right]_{pp}\;.
\label{eq:epsilon1}
\end{equation}
The off-diagonal elements determine the response of
the expansion coefficients for $p\neq q$
\begin{equation}
U_{pq}^{(1)}=\dfrac{(-C^{(0)\dagger}H^{(1)}C^{(0)})_{pq}}{(\varepsilon_{p}-\epsilon_{q})} \; . 
\label{eq:cpscf-LCAO}
\end{equation}
The diagonal elements of $U^{(1)}$ are zero for the electrical field  perturbation by using the orthogonality relation
\begin{equation}
U^{(1)}_{pp}= 0  \;.
\end{equation}

It is clearly shown that this step needs the matrix 
multiplications with dense eigenfunction coefficients~$C^{(0)}$, which results in a scaling of $O(N^3)$. In real numerical evaluation, the scaling exponents can be fitted using the polynomial scaling expression $t=cN^{\alpha}$ for the CPU time as function of the total number of atoms N, such scaling exponents of first order density matrix  calculation in our previous tests were $\alpha=2.8$ for the atomic displacement perturbation\cite{Shang2017}, and $\alpha=2.5$ for the electric field perturbation~\cite{Shang2018}, which were close to the  $O(N^3)$ scaling.

\subsection{The $O(N)$ method to get the first order density matrix}
\label{subsec:ON}
In order to reduce the $O(N^3)$ scaling of the last section, the multiplications with the dense eigenfunction coefficients~$C^{(0)}$ need to be avoided, and the purification related method~\cite{Niklasson2002PRB,Niklasson2003JCP,Niklasson2004,Weber2004}  is a promising choice.  Here we focus on the 
orthogonal formulation of the second order trace-correcting
purification (TC2) method proposed by Niklasson~{\it et al.}~\cite{Niklasson2002PRB,Niklasson2007},
which is a very efficient\cite{Rudberg2011} density-matrix-based method for linear scaling electronic structure calculations. The TC2 method is also
called the second-order spectral projection (SP2) method~\cite{Bock2018} with the same algorithm. 

The TC2 method is initially proposed\cite{Niklasson2002PRB} to solve the KS eigenvalue problem (Eq. \ref{eq:KS-matrix}), which allows us to obtain the density matrix $P^{(0)}$ from ground state Hamiltonian matrix
$H^{(0)}$ directly without the need of performing a matrix diagonalization. It is based on a recursive expansion of the Fermi operator.  The density matrix in atomic basis set is defined as 
\begin{equation}
P=\sum_{i}^{N_{occ}}C_iC_i^{\dagger} \;.
\end{equation}
where $N_{occ}$ is the number of occupied states. And we can get the idempotency relation in the non-orthogonal form
\begin{equation}
PSP = P \;,
\end{equation}
by using Eq.~\ref{eq:orthonormal}.

In order to have the idempotency relation in the orthogonal form,
we firstly transform the Hamiltonian matrix
$H^{(0)}$ to its orthogonal representation~($H^{(0)}_{orth}$) using 
L\"{o}wdin orthogonalization~\cite{Lowdin1950,Lowdin1956}
\begin{equation}
H^{(0)}_{orth} = S^{(0)-\frac{1}{2}} H^{(0)} S^{(0)- \frac{1}{2}}  \;, 
\end{equation}
\begin{equation}
C^{(0)}_{orth} = S^{(0)\frac{1}{2}} C^{(0)}  \;.
\end{equation}
It should be noted that the square root of the overlap matrix needed in
the above L\"{o}wdin orthogonalization is also calculated with 
the linear scaling algorithm~\cite{Jansik2007}  based on the Newton-Schulz iterations. Then we have the orthogonal form of the KS equation,
\begin{equation}
H_{orth}C_{orth}= \epsilon C_{orth}  \;.
\end{equation}
And finally we have the orthogonal form of the density matrix
\begin{equation}
P^{(0)}_{orth} =  S^{(0)\frac{1}{2}} P^{(0)} S^{(0)\frac{1}{2}} \;,
\end{equation}
with the idempotency relation in the orthogonal form,
\begin{equation}
P_{orth}^{(0)}P_{orth}^{(0)} = P_{orth}^{(0)}  \;,
\end{equation}
And this is the base for the orthogonal TC2 method. The initial matrices $X_{0}^{(0)}$ can be written as
\begin{equation}
X_{0}^{(0)}=\dfrac{\epsilon_{max}-H_{orth}^{(0)} }{\epsilon_{max}-\epsilon_{min}}   \;,
\end{equation}
whereby the $\epsilon_{min}$ and $\epsilon_{max}$ denote the minimal and the maximum boundary for the eigenvalues of Hamiltonian matrix $H^{(0)}$, which is estimated with Gershgorin’s Circle Theorem to avoid solving the eigenvalue problem. We use $X_{n}^{(0)}$ to represent the intermediates form of the  $P^{(0)}_{orth}$, and we have the TC2 main cycles: 
\begin{equation}
X_{n+1}^{(0)}= \left\{ \begin{array}{ll}
(X_{n}^{(0)})^2 & \textrm{ $Tr(X_{n}^{(0)})\geq N_{occ}$ }\\
2X_{n}^{(0)}-(X_{n}^{(0)})^2& \textrm{ $Tr(X_{n}^{(0)})< N_{occ}$}
\end{array} \right.
\label{eq:TC2}
\end{equation}
Finally the zero order orthogonal density matrix is gotten after the TC2 cycles are converged:
\begin{equation}
P^{(0)}_{orth}=\lim_{n\rightarrow\infty} {X_{n}^{(0)}} \;.
\end{equation}
We can transform it back to the non-orthogonal density matrix as
\begin{equation}
P^{(0)}=S^{(0)-\frac{1}{2}} P^{(0)}_{orth}  S^{(0)-\frac{1}{2}}  \;.
\end{equation}
%The pseudocode for the TC2 algorithm for DFT is given in Algorithm~\ref{algo:TC2}. 
%
%\begin{algorithm}[H]
%\SetAlgoLined
%\emph{$\mathrm{subroutine}$  $\mathrm{TC2}(H,\rho,N_e)$} \\
%\KwIn{$H^{(0)}$,$S^{(0)}$}
%\KwOut{$P^{(0)}$ }
%H$^{(0)}_{orth}$ = S$^{(0)-\frac{1}{2}}$ H$^{(0)}$ S$^{(0)-\frac{1}{2}}$ \\
%estimate $\varepsilon_{min}(H^{(0)}_{orth}),\varepsilon_{max}(H^{(0)}_{orth})$ \\
%$X_{0}^{(0)}=(\varepsilon_{max}I- H^{(0)}_{orth})/(\varepsilon_{max}-\varepsilon_{min})$ \\
%\For{$n=0; n \le max; n++$} 
%{
% 
%\eIf{$Tr[X_{n}^{(0)}]-N_{occ} <0$}{
% $X_{n+1}^{(0)}=2X_{n}^{(0)}-(X_{n}^{(0)})^{2}$\\
% }{
%   $X_{n+1}^{(0)} = X{n}^{(0)})^2$ 
% }
%  
%\emph{estimate Error}
%
%\If{$Error < Error Limit$} { break }
% } 
%P$^{(0)}$ = S$^{(0)-\frac{1}{2}}$ X$_{n}^{(0)}$ S$^{(0)-\frac{1}{2}}$ \\
% \caption{The TC2 algorithm, where $N_{occ}$ is the number of occupied states, $\varepsilon_{min}(H)$/$\varepsilon_{max}(H)$ denotes the minimal/maximum
% boundary for the eigenvalues of Hamiltonian matrix $H^{(0)}$, which is estimated with Gershgorin’s Circle Theorem.}
%\label{algo:TC2}
%\end{algorithm}

\begin{figure}
 \centering
 \includegraphics[width=0.9\columnwidth]{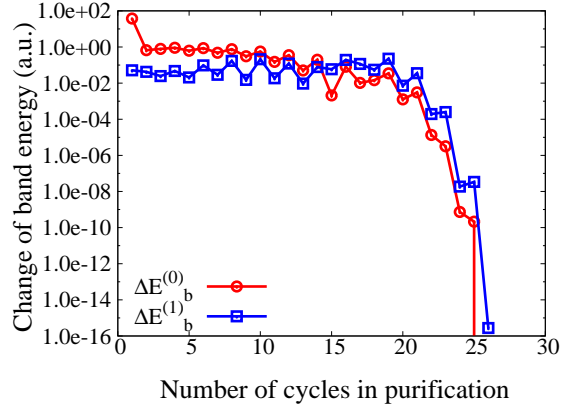}
 \caption{The change of zero band energy~($E^{(0)}_{b}=\rm{Tr}[P^{(0)}H^{(0)}]$) and the first order band energy~($E^{(1)}_{b}=\rm{Tr}[P^{(1)}H^{(0)}]$) with respect to number of recursive cycles in Eq.~(\ref{eq:TC2})  and Eq.~(\ref{eq:TC2-CPSCF}) respectively. Here the NTPoly-filter is set to $10^{-8}$ and NTPoly-tolerance is set to $10^{-4}$.  }
\label{fig: band_energy}
 \end{figure}

Such TC2 method can be extended to the response theory directly~\cite{Weber2004,Niklasson2004}, which provides explicit construction of the derivative density matrix, i.e. another way to get the first order 
density matrix directly from the first order Hamiltonian.  Here we first
define the initial first order matrices $X_{0}^{(1)}$ as
\begin{equation}
X_{0}^{(1)}=\dfrac{-H^{(1)} }{\epsilon_{max}-\epsilon_{min}}
\end{equation}
where $\epsilon_{max}$ and $\epsilon_{min}$ are the maximal and minimal eigenvalues of the unperturbed Hamiltonian $H^{(0)}$, then using the following recursive cycles in Eq.(\ref{eq:TC2-CPSCF}), we get the first order density matrix per CPSCF cycle: 
\begin{equation}
X_{n+1}^{(1)}= \left\{ \begin{array}{ll}
X_{n}^{(1)}X_{n}^{(0)}+X_{n}^{(0)}X_{n}^{(1)}& \textrm{ $Tr(X_{n}^{(0)})\geq N_{occ}$ }\\
2X_{n}^{(1)}-X_{n}^{(1)}X_{n}^{(0)}-X_{n}^{(0)}X_{n}^{(1)}& \textrm{ $Tr(X_{n}^{(0)})< N_{occ}$}
\end{array} \right.
\label{eq:TC2-CPSCF}
\end{equation}
Finally the first order orthogonal density matrix is gotten after the recursive cycles are converged:
\begin{equation}
P^{(1)}_{orth}=\lim_{n\rightarrow\infty} {X_{n}^{(1)}} \;.
\end{equation}
We then transform it back to the non-orthogonal first order density matrix as
\begin{equation}
P^{(1)}=S^{(0)-\frac{1}{2}} P^{(1)}_{orth}  S^{(0)-\frac{1}{2}}  \;.
\end{equation}
These equations provide the base for computing the density-matrix
response explicitly and rapidly. 
%The orthogonal form of such  
%TC2-CPSCF scheme is described by Algorithm
%\ref{algo:TC2-CPSCF}. 
In this work, such TC2 recursive algorithm for the first order density matrix has been combined with the CPSCF cycles, and implemented in the all-electron Fritz Haber Institute \textit{ab initio} molecular simulations (FHI-aims) package~\cite{Ren/etal:2012,Havu/etal:2009}. As shown in Fig.\ref{fig: band_energy}, the change of the zero order band energy~($E^{(0)}_{b}=\rm{Tr}(P^{(0)}H^{(0)})$) and the first order band energy($E^{(1)}_{b}=\rm{Tr}(P^{(1)}H^{(0)})$) converged fast with respect to the number of the recursive cycles, after around 25 cycles, the change of the band energy is reduced to $10^{-10}$ a.u..

%\begin{algorithm}
%\SetAlgoLined
%\emph{$\mathrm{subroutine}$  $\mathrm{TC2-CPSCF}(H,\rho,N_e)$} \\
%\KwIn{H$^{(0)}$,S$^{(0)}$,H$^{(1)}$}
%\KwOut{$P^{(0)}$, P$^{(1)}$}
%H$^{(0)}_{orth}$ = S$^{(0)-\frac{1}{2}}$ H$^{(0)}$ S$^{(0)-\frac{1}{2}}$ \\
%H$^{(1)}_{orth}$ = S$^{(0)-\frac{1}{2}}$ H$^{(1)}$ S$^{(0)-\frac{1}{2}}$  \\
%estimate $\varepsilon_{min}(H^{(0)}_{orth}),\varepsilon_{max}(H^{(0)}_{orth})$ \\
%$X_{0}^{(0)}=(\varepsilon_{max}I- H^{(0)}_{orth})/(\varepsilon_{max}-\varepsilon_{min})$ \\
%$X_{0}^{(1)}=(- H^{(1)}_{orth})/(\varepsilon_{max}-\varepsilon_{min})$ \\
%\For{$n=0; n \le max; n++$} 
%{
% \eIf{$Tr[X_{n}^{(0)}]-N_{occ} <0$}{
% $X_{n+1}^{(0)}=2X_{n}^{(0)}-(X_{n}^{(0)})^{2}$\\
%$  X_{n+1}^{(1)} =2X_{n}^{(1)}-X_{n}^{(1)}X_{n}^{(0)}-X_{n}^{(0)}X_{n}^{(1)}$
% }{
%   $X_{n+1}^{(0)} = X{n}^{(0)})^2$ \\
%  $ X_{n+1}^{(1)} = X_{n}^{(1)}X_{n}^{(0)}+X_{n}^{(0)}X_{n}^{(1)}$ 
% }
%  \emph{estimate Error}
%\If{$Error < Error Limit$} { break }
% } 
%P$^{(0)}$ = S$^{(0)-\frac{1}{2}}$ X$_{n}^{(0)}$ S$^{(0)-\frac{1}{2}}$ \\
%P$^{(1)}$ = S$^{(0)-\frac{1}{2}}$ X$_{n}^{(1)}$ S$^{(0)-\frac{1}{2}}$  \\
% \caption{The TC2-CPSCF algorithm, where $N_{occ}$ is the number of occupied states,  $\varepsilon_{min}(H)$/$\varepsilon_{max}(H)$ denotes the minimal/maximum boundary for the eigenvalues of Hamiltonian matrix $H^{(0)}$, which is estimated with Gershgorin’s Circle Theorem.}
%\label{algo:TC2-CPSCF}
%\end{algorithm}

\subsection{The parallel algorithm for sparse matrix multiplication}
\label{sec:spdgemm}
\begin{figure}
 \centering 
 \includegraphics[width=0.95\columnwidth]{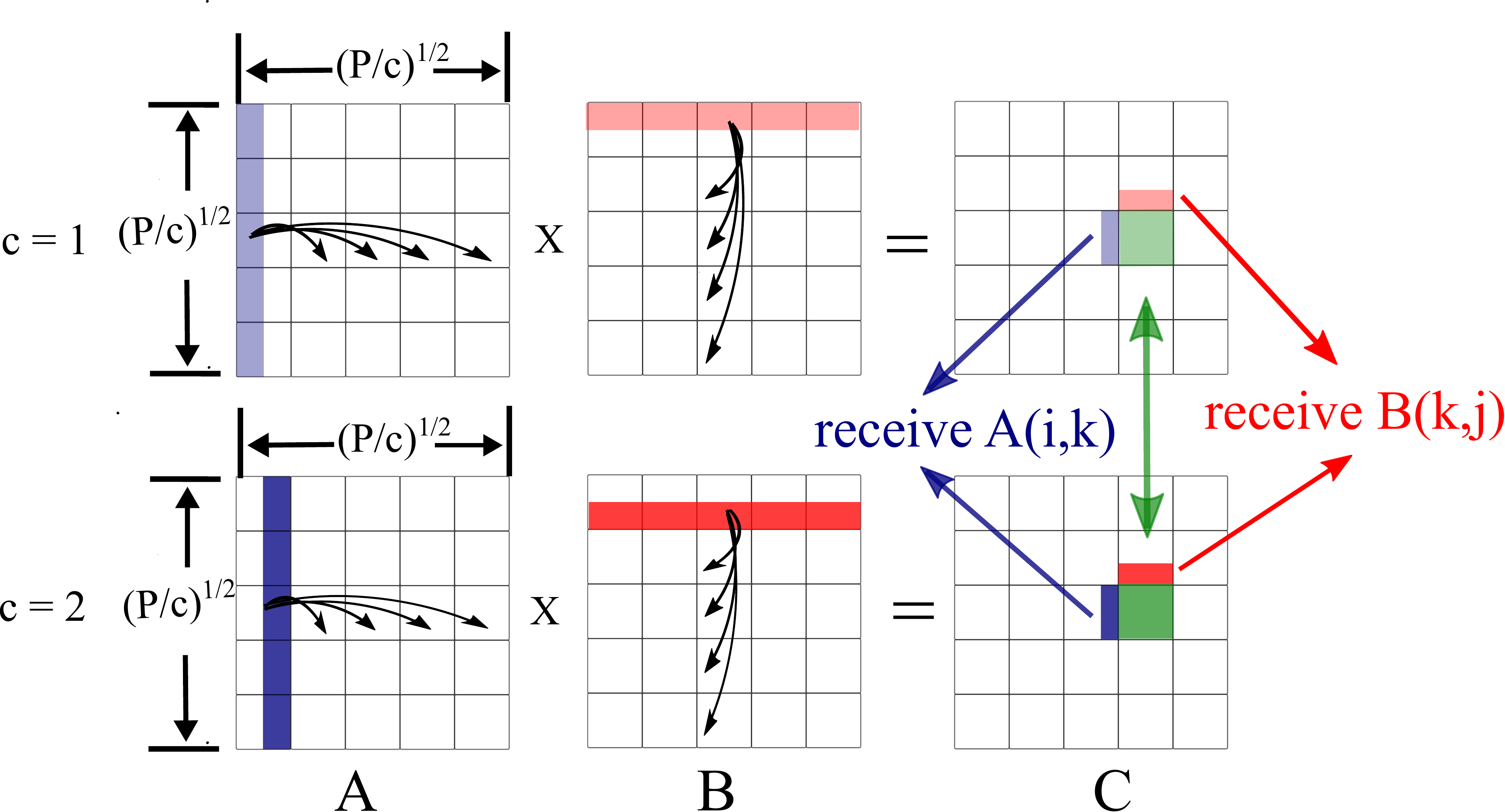}
 \caption{The 3D-SpGEMM algorithm for spare matrix-matrix multiplication $C=A \times B$ on a $\sqrt{P/c} \times \sqrt{P/c} \times c$ processor grid. Here c in the Z direction is set to 2 in this illustration.  The matrix elements of A and B are broadcasted and multiplied locally to compute a contribution to a local result matrix C, and then C matrix are merged across the Z direction to get the final results. }
 \label{fig:3D-SpGEMM}
 \end{figure} 
 
%%---------------begin the version for \usepackage[ruled]{algorithm2e}------------
%\begin{algorithm}
%\SetAlgoLined
%\emph{$\mathrm{subroutine}$ $\mathrm{ROWWISE-SpGEMM}$ } \\
%\KwIn{A, B}
%\KwOut{C}
%\For{$i=1; i \le n; i++$} 
%{
%  \For{$k=1; k \le n; k++$}
%  {
%  \If{ \rm{A(i,k)}$ \neq 0 $} {
%   C(i,:) = C(i,:) + A(i,k) B(k,:)   \\ } 
% 
%  }
% } 
% \caption{The row-wise formulation of the serial Gustavson's sparse matrix multiplication for compressed sparse  rows~(CSR) data format.}
%\label{algo:CSR-SpGEMM}
%\end{algorithm}
%%--------------end the version for \usepackage[ruled]{algorithm2e}------------

\begin{algorithm}
\caption{The serial sparse matrix-matrix multiplication algorithm with the compressed sparse rows~(CSR) data format. The CSR representation of a sparse matrix A is given by three one-dimensional array IA, JA, and A. IA is the (address) pointer of the first nonzero element for the rows of A;  JA is the column indices of the nonzero matrix elements; A is the numerical values of the nonzero matrix elements.  }
\begin{algorithmic}
    \Require \\
    nrow : the row dimension of A and C   \\
    ncol : the column dimension of B and C  \\
    IA, JA, A:  input sparse matrix A with CSR format  \\
    IB, JB, B:  input sparse matrix B with CSR format  \\
    IC, JC, C:  output sparse matrix  C with CSR format    
\end{algorithmic}  

\begin{algorithmic}
\State IC(1) $\gets$ 1 
\State C $\gets$ 0 
\For{ k $\gets$ 1, ncol  }
\State label(k) $\gets$ 0 
\EndFor 
\For{i$\gets$ 1, nrow}
          \For{pA  $\gets$ IA(i), IA(i+1)-1}
          \State j $\gets$ JA(pA)       
             \For{pB $\gets$ IB(j), IB(j+1)-1  }
             \State k $\gets$ JB(pB)
                \If{label(k) .eq. 0}               
                  \State  pC $\gets$ pC+1  
                  \State  JC(pC) $\gets$ k     
                  \State  label(k) $\gets$ pC 
                  \State  C(pC) $\gets$ A(pA) $\times$ B(pB) 
                \Else
                  \State  C(label(k)) $\gets$ C(label(k)) + A(pA) $\times$ B(pB) 
                \EndIf                
            \EndFor   % pB
         \EndFor      % pA 
         \For{ pCN $\gets$ IC(i), pC  }                       
            \State  label(JC(pCN)) $\gets$ 0  
         \EndFor
         \State IC(i+1) = pC + 1     
\EndFor   % i = 1, nrow
\end{algorithmic} 
\label{algo:CSR-SpGEMM}
\end{algorithm}

The parallel performance bottleneck in the above TC2 and TC2-CPSCF
methods is the sparse matrix-matrix multiplication. Here in this subsection
we will show how this sparse matrix-matrix multiplication is performed.  
The serial algorithm for sparse matrix-matrix multiplication in compressed sparse  rows~(CSR) format is given by
Gustavson\cite{Gustavson1978}, as shown in Algorithm~\ref{algo:CSR-SpGEMM}. 
In its parallelization, the so-called 3D-SpGEMM algorithm developed by Ballard ~{\it et al.}~\cite{Ballard2013}  and Adaz~{\it et al.}~\cite{Azad2016} is employed to minimize data communication between processors in the parallel progress, which effectively optimizes the parallel computations of the sparse matrix-matrix multiplication. In this algorithm, as shown in Fig.~\ref{fig:3D-SpGEMM}, each matrix is distributed along the cubic $\sqrt{P/c}\times \sqrt{P/c} \times c$ processor grid, where $1<c<\sqrt[3]{P}$ and $P$ is the total number of the processors. Then each matrix is broadcasted and multiplied locally to compute a contribution to a local result matrix, and finally the result matrix is summed up.

This 3D-SpGEMM algorithm has been implemented in the Combinatorial BLAS library\cite{Bulu2011} as well as in the NTPoly library~\cite{Dawson2018}, a library for massively parallel sparse matrix function calculations. The algorithm shows very good strong
scaling performance for sparse matrix multiplications\cite{Dawson2018}. Such NTPoly package has been 
integrated into ELSI\cite{YU2018}, which is a general open-source infrastructure 
for large-scale electronic structure theory and can be linked with FHI-aims and SIESTA. 
Here our implementation of the TC2-CPSCF method is  based on the sparse matrix-matrix multiplication routine from NTPoly in ELSI. The sparse matrix is stored with compressed sparse row~(CSR) storage format in FHI-aims. In NTPoly, the coordinate
format~(COO) data format is adopted as the input interface. 
In the COO data format, the global row, column and value are stored in a triplet list, which is convenient to make parallel decomposition of the global parallel matrices into local sparse matrices. Then the local sparse matrix stored in the COO format is transformed to the CSR format to perform the local matrix-matrix multiplication with Algorithm~\ref{algo:CSR-SpGEMM}.  In order to use NTPoly in FHI-aims, we first need to translate the CSR storage format to the triplet format,
and then performs the TC2/TC2-CPSCF scheme to get the density matrix/first order density matrix, finally the data is transformed back to the CSR format in FHI-aims. 
It should be noted that since the input sparse matrices in our calculations are usually not well distributed in parallel processes, the rows and columns of the matrix need to be randomly permuted by multiplying the sparse matrices with permutation matrices, in order to achieve the load balance in the sparse matrix-matrix multiplication.

In the NTPoly implementation, two parameters are used to control
the error. One is called NTPoly-filter, which refers to the threshold to determine which matrix elements can be treated as zero. This parameter scales linearly with the accumulated density matrix error. 
It should be noted that the accumulated error of the purification method is bounded and related to the drop tolerance of matrix elements and the band gap of the system~\cite{Niklasson2003JCP}, but it is difficult to be controlled with rigorous numerics\cite{Rubensson2005,Rubensson2008}. The strategy to rigorously control the forward error of density matrix purification can be found in Ref.~\cite{Rubensson2008}. 
The other parameter is called NTPoly-tolerance, which is the convergence-threshold 
which compared the band energy between the current iteration and the last iteration.  In the following, we will give the examination for the two parameters in real applications.

\section{Results}
\label{sec:result}
To validate our implementation we have specifically investigated the convergence of polarizabilities with respect to the numerical parameters used in the TC2-CPSCF calculation in Sec.~\ref{subsec:convergence}. Furthermore, a systematic validation of the TC2-CPSCF implementation by comparing to polarizabilities obtained from the benchmark $O(N^3)$ method is presented in Sec.~\ref{subsec:validation}. The computational performance of the TC2-CPSCF implementation is discussed in Sec.~\ref{subsec:performance}. 

In FHI-aims, the atom-centered integration grids are used for the numerical integration. Each atom has the radial shells around it, and the angular points are distributed on each radial shell. The grid settings in FHI-aims are described by light, tight and really-tight with different radial shells and angular integration points, the tighter the better quality of the integration grid. The basis set setting in FHI-aims are defined as following: A minimal basis includes the radial functions of the occupied orbitals of free atoms with noble gas configuration, then the quantum numbers of the additional valence functions,  and additional radial functions are added to make ``tier 1'' ,``tier 2'', and so on. Such basis sets are similar to the split-valence polarization basis used in the Gaussian basis set. For example, the ``tier 1'' basis set is equivalent to the double-zeta plus polarization basis set.
The parameter c discussed in Sec.~\ref{sec:spdgemm} is set to 1 in the following calculations, because the c$>$1 setting only shows better performance when the number of the CPU cores is larger than 10,000, as shown in Ref.\cite{Dawson2018}. 

\subsection{Convergence with respect to numerical parameters}
\label{subsec:convergence}
In this part, the convergence behaviour of the TC2-CPSCF method with respect to the numerical parameters~(NTPoly-filter, NTPoly-tolerance) is analysed. We use the water~(H$_2$O) molecule as an example, for which we 
compute the three diagonal components of the polarizability tensor using a local approximation for exchange and correlation (LDA parametrization of Perdew and Zunger~\cite{Perdew/Zunger:1981} for the correlation energy density of the homogeneous electron gas based on the data of Ceperley and Alder~\cite{Ceperley/Alder:1980}). The tight setting is adopted for the integration and the ``tier 2'' basis set is adopted in this calculation. 

\begin{figure}
 \centering 
\includegraphics[width=0.7\columnwidth]{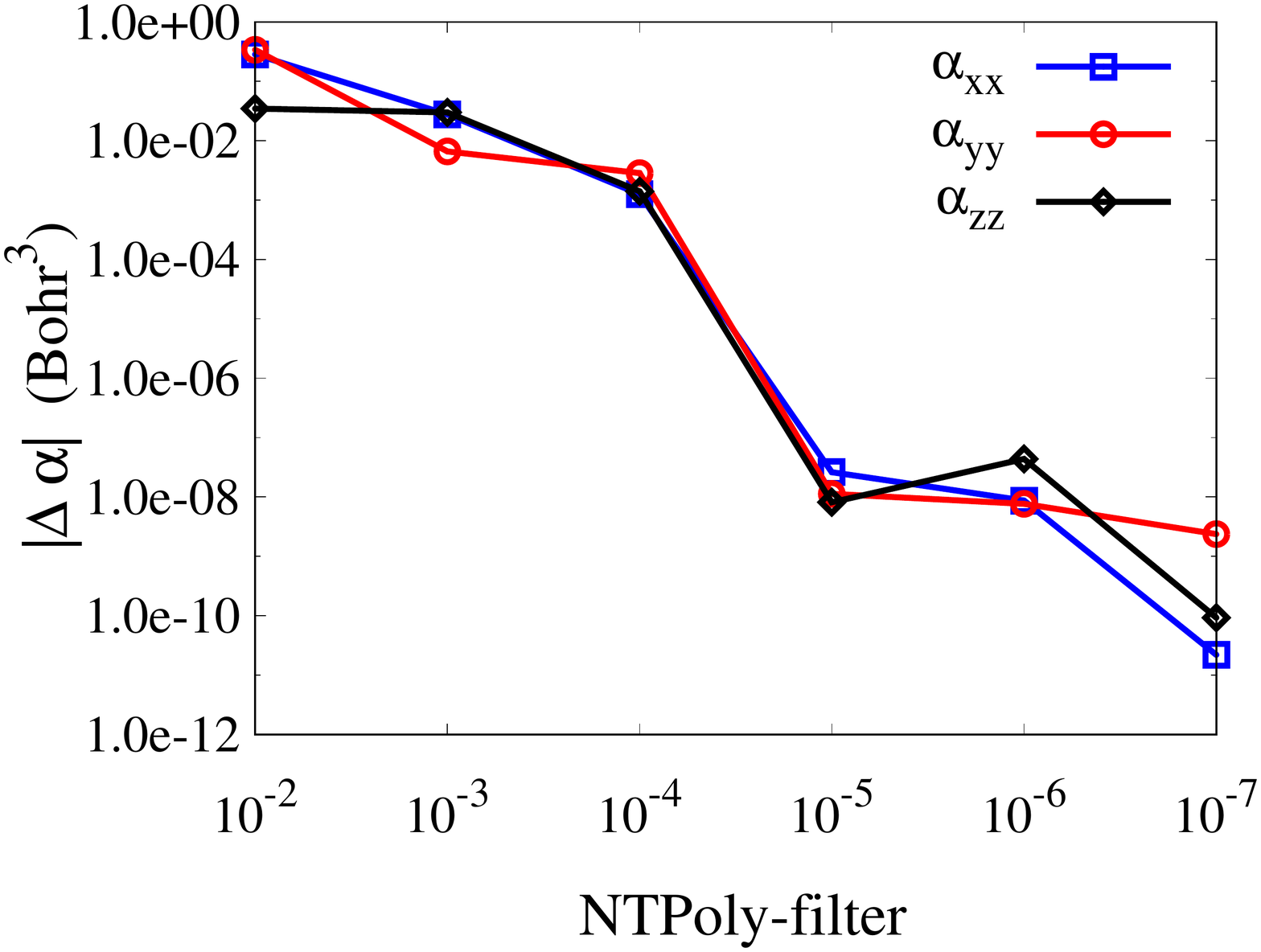}   \\
\includegraphics[width=0.7\columnwidth]{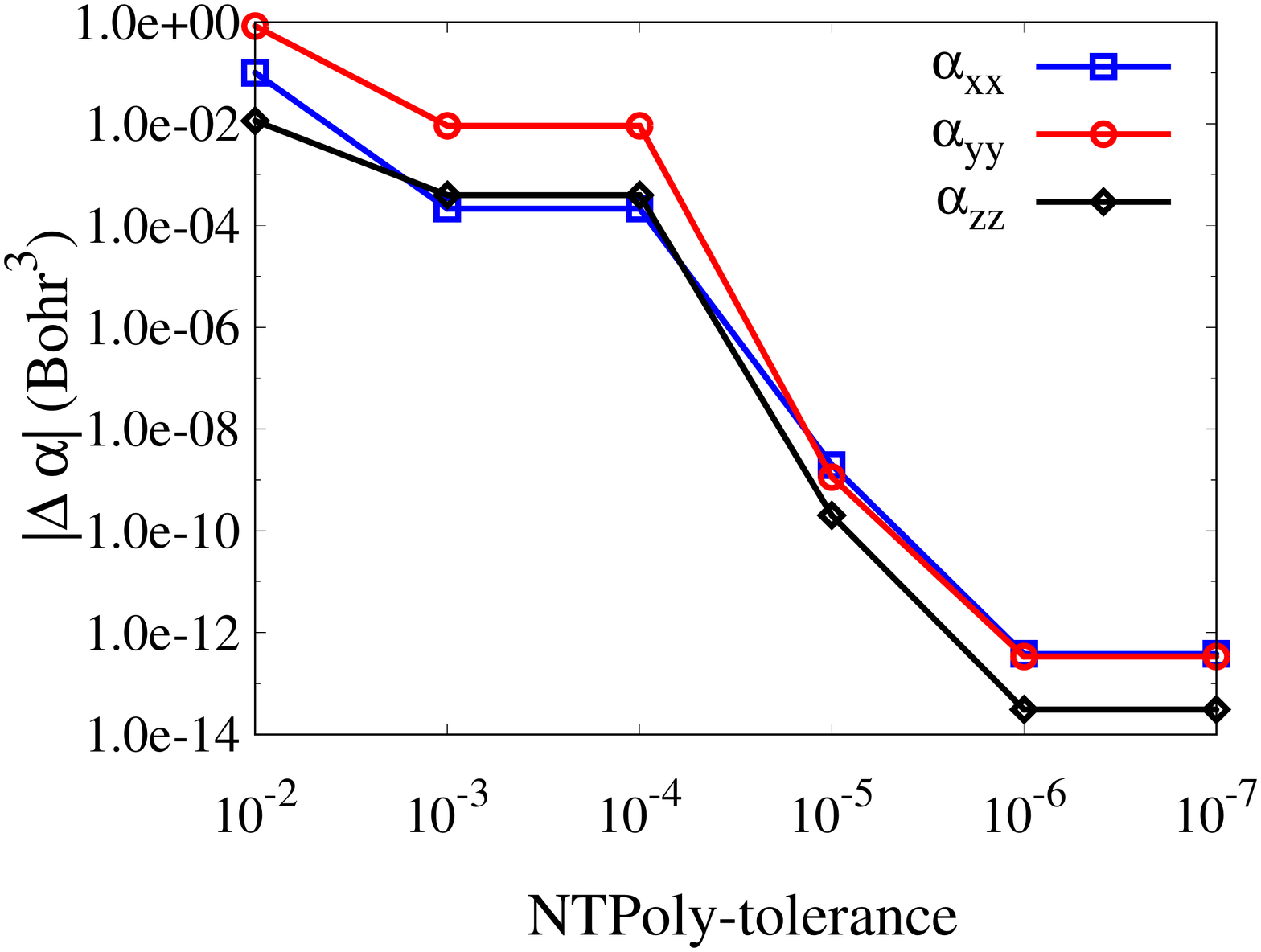} 
 \caption{The convergence of the absolute error of three diagonal components of the polarizability tensor of H$_2$O with respect to NTPoly-filter and NTPoly-tolerance. Here in the upper panel, the NTPoly-tolerance is fixed to $10^{-4}$ while the NTPoly-filter is changed from
$10^{-2}$ to $10^{-7}$; In the lower panel, the NTPoly-filter is fixed to $10^{-8}$ while the NTPoly-tolerance is changed from $10^{-2}$ to $10^{-7}$. }
 \label{fig:h2o_convergence}
 \end{figure}

The upper panel of Fig.\ref{fig:h2o_convergence} shows the absolute error change in the three diagonal components of the polarizability if the NTPoly-filter is changed. The NTPoly-filter is the parameter to 
determine the threshold smaller than which the matrix elements will be discarded in the process of zero/first order density matrix purification. Here, the NTPoly-filter is changed from
$10^{-2}$ to $10^{-8}$ and the NTPoly-tolerance is fixed to $10^{-4}$. We can see the polarizabilities converged quickly 
with respect to the NTPoly-filter. At around NTPoly-filter=$10^{-4}$, we get the  maximal absolute/relative error of 0.002 Bohr$^3$/0.06\% with respect to the NTPoly-filter=$10^{-8}$ setting. 
  
The lower panel of Fig.\ref{fig:h2o_convergence} shows the convergence
test with respect to NTPoly-tolerance, which is the parameter to
determine the convergence criterion of the zero/first order density matrix purification. We change the  NTPoly-tolerance from $10^{-2}$ to $10^{-8}$ and fixed the NTPoly-filter to $10^{-8}$, and the polarizabilities converged also fast with respect to the NTPoly-tolerance. At NTPoly-tolerance=$10^{-3}$, we get the maximal absolute/relative error of 0.009 Bohr$^3$/0.2\% with respect to the NTPoly-tolerance=$10^{-8}$ setting. 
  
As a result, in the following calculation, we can safely use NTPoly-filter~($10^{-8}$) and NTPoly-tolerance~($10^{-4}$) in the validation part in Sec.~\ref{subsec:validation}. Moreover, it is also enough for us to use NTPoly-filter~($10^{-6}$) and NTPoly-tolerance~($10^{-5}$) in the performance evaluation part in Sec.~\ref{subsec:performance}.   

\subsection{Validation against benchmark results}
\label{subsec:validation} 
The polarizabilities of 32 selected molecules are 
calculated with the linear scaling TC2-CPSCF method described in Sec.~{\ref{subsec:ON}. The results are compared with the normal $O(N^3)$ method described in Sec.~{\ref{subsec:ON3} to serve as the benchmark to make the validation.  The detailed comparison for each individual molecule is listed in the~\ref{appendix:mol}. Here we summarized the data in Tab.~\ref{tab:mae},  where we list the mean absolute percentage error~(MAPE) and the mean absolute error~(MAE) for all tested molecules. Overall, we find an excellent agreement between our $O(N)$ TC2-CPSCF method and the $O(N^3)$ benchmark results. 

\begin{table}[h]
\centering
\begin{tabular}{c |  c c }
\hline \hline
         &   MAE~(Bohr$^{3}$) &  MAPE \\
\hline 
Dimers &    0.023  & 0.078\%   \\ 
Molecules & 0.0036 & 0.015\% \\
\hline \hline
\end{tabular}
\caption{
Mean absolute error (MAE) and mean absolute percentage error (MAPE) for the difference between the polarizabilities obtained via linear scaling TC2-DFPT method for a set of 16 dimers, 16 molecules. All calculations are performed at the LDA level of theory with fully converged
numerical settings and relaxed geometries. Detailed informations including the values for each individual molecule can be found in the Appendix.}
\label{tab:mae}
\end{table}

\subsection{Performance of the implementation}
\label{subsec:performance}
To demonstrate the scaling performance of our implementation,
we use the H(C$_2$H$_4$)$_n$H molecules oriented along the X-axis as shown in Fig.~\ref{fig:c2h4-mol} as the examples. All calculations use light settings and the LDA functional. Calculations were performed on three node of  Intel(R) Xeon(R) CPU E5-2678v3 CPUs~(12 cores each at 2.50GHz).
%this is bingxing-9. 

\begin{figure}
\includegraphics[width=1.0\columnwidth]{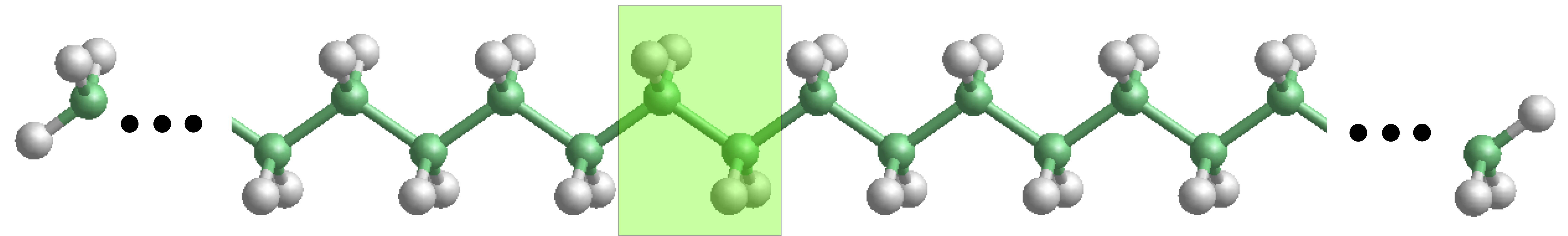}
\caption{The H(C$_2$H$_4$)$_n$H molecules used in this work to do the performance test. Here the repeated unit is the C$_2$H$_4$ block marked with a green shade, the number n is chosen to change from 64 to 640, which corresponding to the number of atoms change from 386 to 3842 in our performance test. The white ball is hydrogen atom and green one is carbon atom. }
\label{fig:c2h4-mol}
\end{figure}

%--------sparsity---------
We first investigate the matrix sparsity and the time scaling with respect to the number of atoms. In our DFPT implementation, the CPSCF is performed for each coordinate independently. As the H(C2H4)$_n$H molecule is placed along X-axis, we just examine the DFPT perturbation for X and Z coordinate respectively, since the Y coordinate gives the same result as the Z coordinate.  In Fig.\ref{fig:sparsity}(a), we can see that both the density matrix sparsity and the first order density matrix decay as $O(\frac{1}{N})$ with respect to the number of atoms. The sparsity of the first order 
density matrix in X-axis is larger than in Z-axis, this is because the electric field in X-axis just polarizes the electric density in this direction and causes the density overlapping since the H(C2H4)$_n$H line was also placed along the X-axis. On the other hand, the sparsity of the first order density matrix in Z-axis is similar with the zero order density matrix because the polarization along Z-axis does not bring the additional overlap of the density. 
In Fig.~\ref{fig:sparsity}(b), the number of the non-zero elements is examined, since the number of the matrix elements increases as $O(N^2)$
and the sparsity increases as $O(\frac{1}{N})$, so the number of the non-zero elements increase as $O(N)$. Since the sparsity of the first order density matrix in X-axis is larger than Z-axis, so the prefactor of the non-zero elements is also larger in X-axis. 
Finally, in Fig.~\ref{fig:sparsity}(c), we show the purification time  
per SCF/CPSCF cycle, in which we can see the DFPT time in Z-axis
is around 3 times of the DFT time, and this is because the number of the matrix operations in DFPT~(TC2-CPSCF) is around 3 times of the DFT~(TC2).
We can also see the DFPT time in X-axis is around 5 times of the DFT because of the sparsity of the response density in X-axis is larger than the one in Z-axis. Finally we observe the overall
linear scaling in both the DFT and DFPT calculations for the purification time with increasing system sizes. 

\begin{figure*}
\subfloat[Sparsity]{\includegraphics[width=0.65\columnwidth]{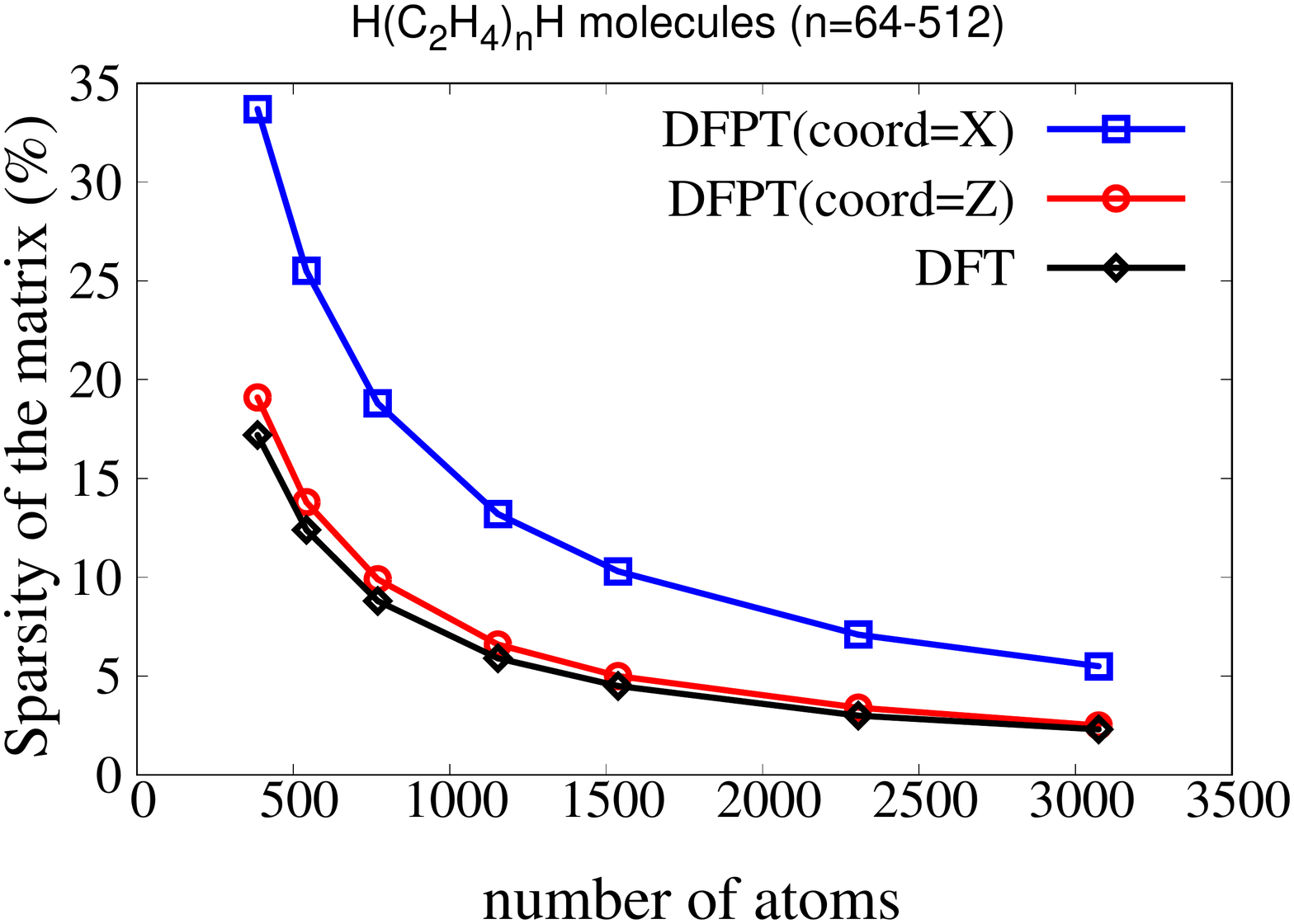} } 
 \subfloat[Non-zero element]{\includegraphics[width=0.65\columnwidth]{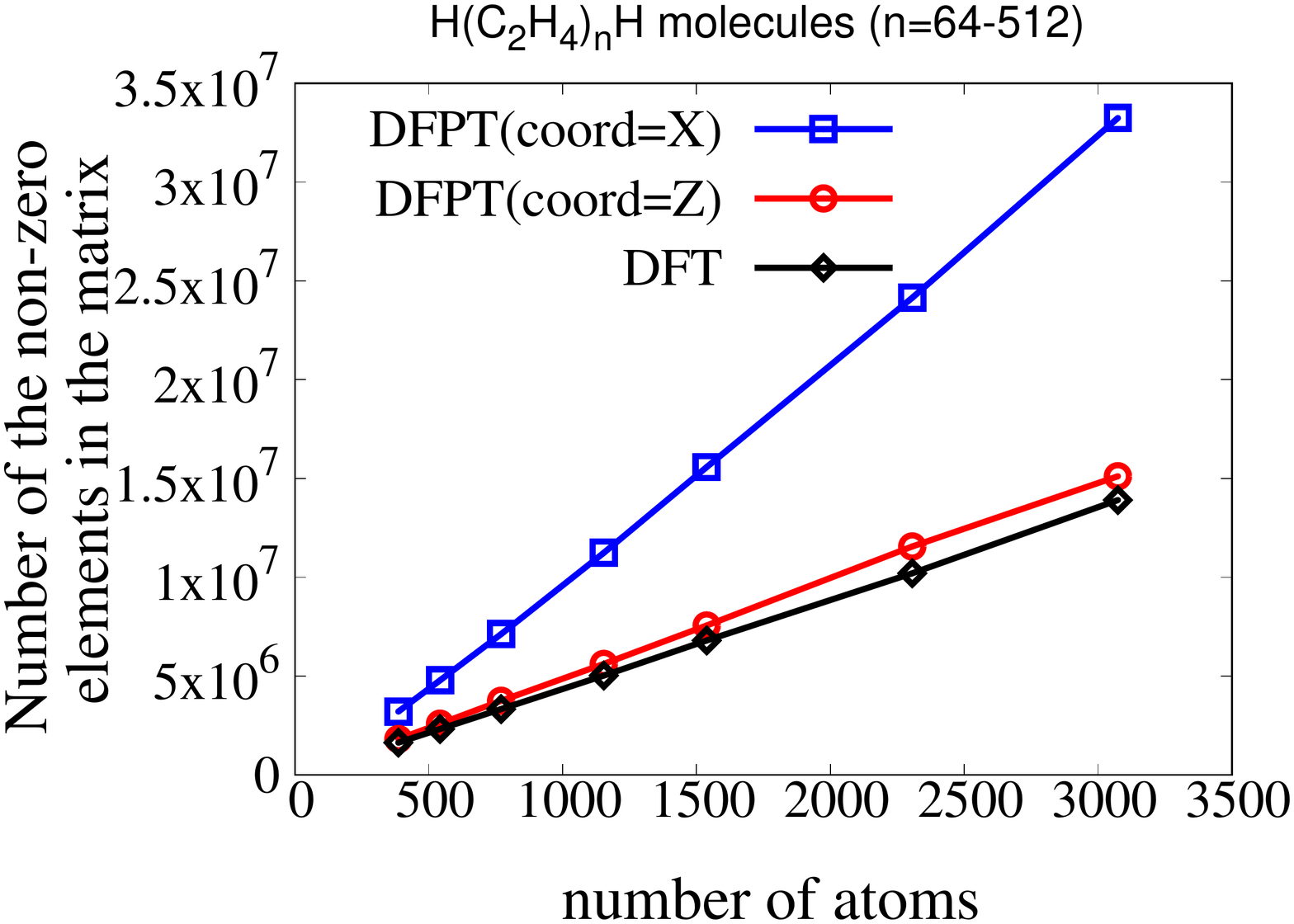} } 
 \subfloat[Time]{\includegraphics[width=0.65\columnwidth]{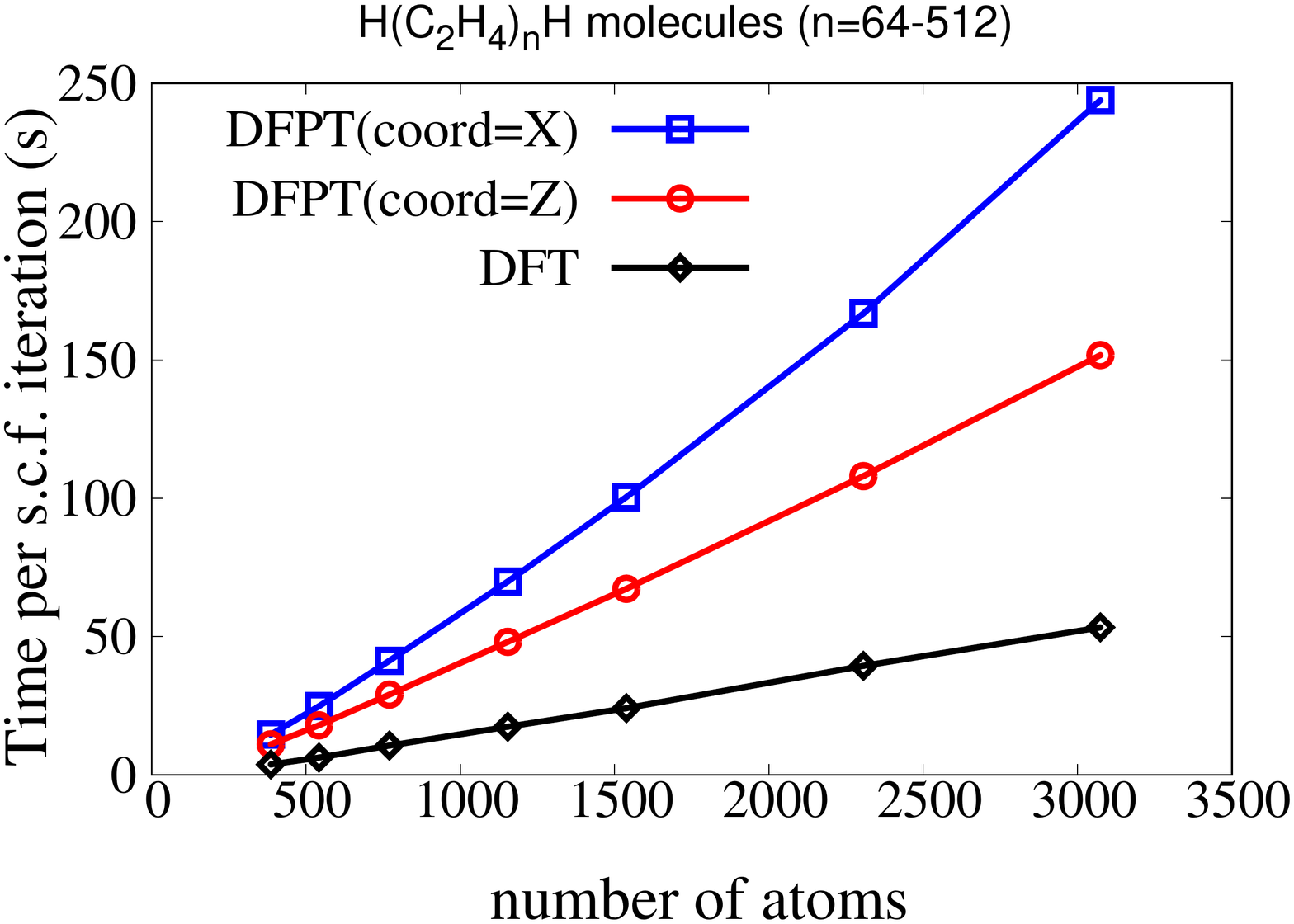}  }
 \caption{The matrix sparsity, number of the non-zero elements and the purification time with respect to the number of the atoms for the DFT and DFPT calculation.  }
 \label{fig:sparsity}
 \end{figure*}

%---------filter------------
In addition to the sparsity, the parameter NTPoly-filter also 
influences the linear scaling prefactor. As shown in Fig.\ref{fig:filter}, two values of NTPoly-filter are adopted, and the computation time with NTPoly-filter~($10^{-6}$) is nearly double of the one computed with NTPoly-filter~($10^{-5}$).

\begin{figure}
 \centering
 \includegraphics[width=0.9\columnwidth]{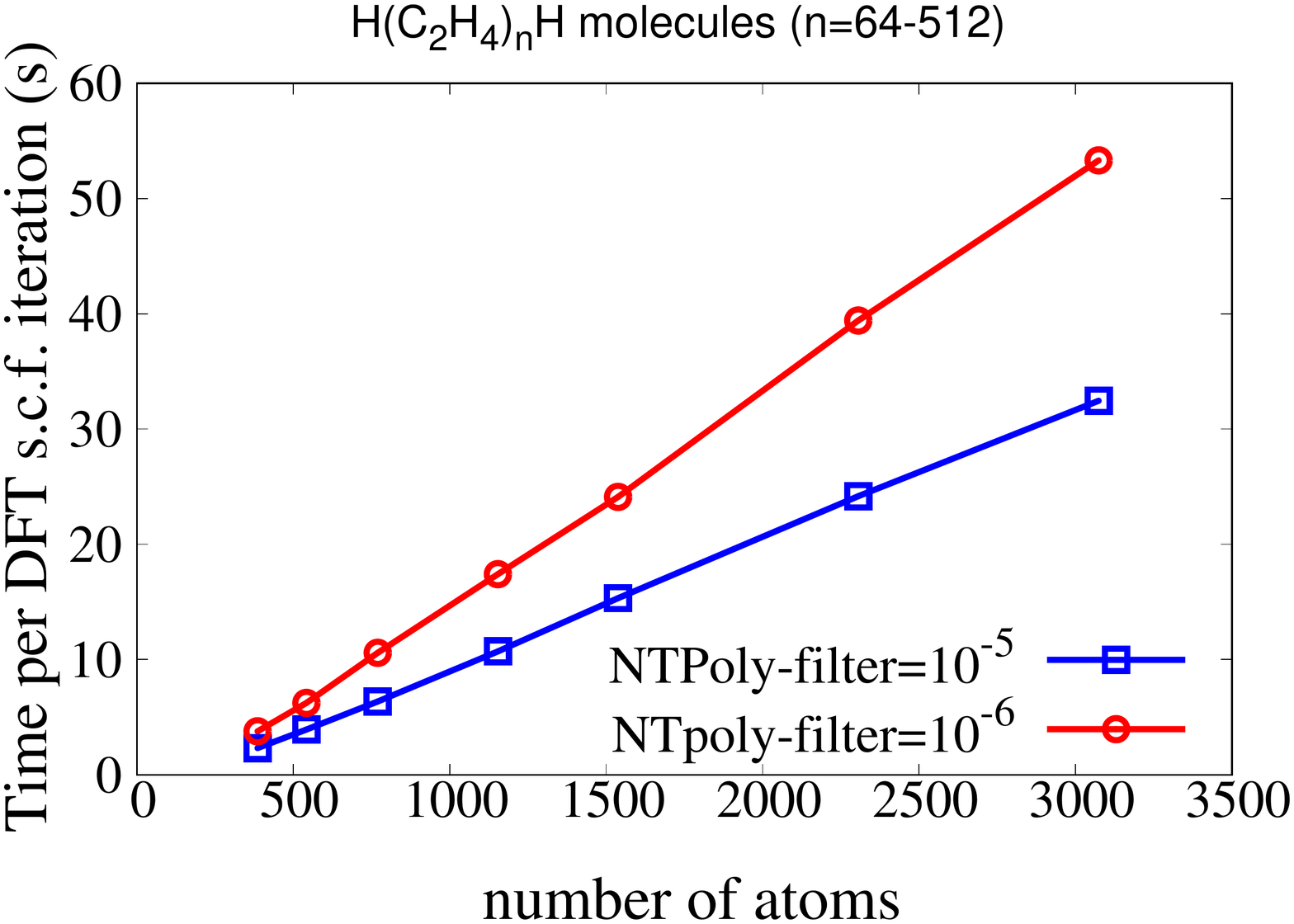}
 \caption{The purification time per DFT iteration with respect
 to the number of the atoms. Here the influence of the NTPoly-filter parameter to the linear scaling prefactor is shown. }
\label{fig:filter}
 \end{figure}

%---------DFT-vs-DFPT--------
In Fig.~\ref{fig:DFT_vs_DFPT}, we present the purification time of the zero/first order density matrix for the isolated H(C2H4)$_n$H molecule
systems with the number of atoms changing from 386 to 3,842~(the corresponding number of the basis functions are changing from 3,082 to 30,730). The DFT-$O(N)$ and DFPT-$O(N)$ mean the TC2 and TC2-CPSCF method respectively as described in Sec.~\ref{subsec:ON}, while the DFT-$O(N^3)$ scaling method refers to the traditional matrix diagonalization algorithm, and the DFPT $O(N^3)$ scaling method refers to the dense matrix algorithm as shown in Sec.\ref{subsec:ON3}. Here the DFPT results are for the perturbation alone the Z direction. It should be noted that, the traditional $O(N^3)$ method is fully optimized both in DFT~\cite{YU2018} and DFPT~\cite{Shang2017,Shang2018} method. The numerical thresholds NTPoly-filter~($10^{-6}$) and NTPoly-tolerance~($10^{-5}$) are applied in the $O(N)$ method. It is clearly shown that the performance
 of the TC2 method is better than the traditional $O(N^3)$  method at around 1300 atoms~(10000 basis functions), and the performance of the TC2-CPSCF method is better than the traditional $O(N^3)$ method at around 3000 atoms~(23000 basis function). The comparison of the total time for the calculation of the polarizabilities between the $O(N)$ TC2-CPSCF method and the traditional $O(N^3)$ method is shown in the~\ref{appendix:total_time}, which gives similar crossover point.  
 
 \begin{figure}
 \centering
 \includegraphics[width=0.9\columnwidth]{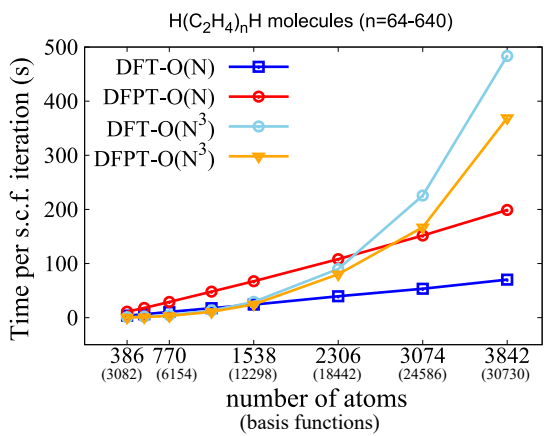}
 \caption{The purification time of the zero/first order density matrix build with DFT/DFPT for isolated H(C2H4)$_n$H molecules containing from 386 to 3842 atoms, with the number of the basis functions changing from 3082 to 30730. Here we use NTPoly-filter~($10^{-6}$) and NTPoly-tolerance~($10^{-5}$) as the numerical thresholds. All calculations are performed on 36 CPU cores. }
 \label{fig:DFT_vs_DFPT}
 \end{figure}

%----------full_cpscf----------
In order to systematically investigate the scaling performance 
of each part in the DFPT calculation, we show in Fig.~\ref{fig:scaling_c2h4_mol} for the CPU time contributed from the individual response properties~(density $n^{(1)}$, electrostatic potential $V^{(1)}$, Hamiltonian matrix $H^{(1)}$, density matrix $P^{(1)}$) as well as the 
total summation of all the contributions~($n^{(1)}$+$V^{(1)}$+$H^{(1)}$+$P^{(1)}$) per DFPT cycle.

The scaling exponents of the computation time (as a function of the total number of atoms N) in calculating each response quantity were fitted using the polynomial scaling formula $t=cN^{\alpha}$($\alpha$ is the exponent), with the obtained exponent values listed in the upper panel of Fig.~\ref{fig:scaling_c2h4_mol}. We find that calculating the first order density matrix $P^{(1)}$ dominates the computational time, which, in principle, exhibits a strict O(N) scaling. For a system whose size is ranged from 386 atoms to 3884 atoms, the obtained exponent $\alpha$ of 1.2 is close to the expected O(N) scaling. Calculating the first order electrostatic response potential $V^{(1)}$ is the second expensive part, and the corresponding exponent $\alpha$ of 1.7 is similar to that in updating the ground-state electrostatic potential~\cite{Blum2009}. For very large systems ($N > 4000$), updating $V^{(1)}$ dominates, since it scales higher than updating $P^{(1)}$. Calculating the Hamiltonian response matrix $H^{(1)}$ and the first order response density $n^{(1)}$ corresponds to an exponent $
\alpha$ of 1.6, since it involves similar numerical operations.

\begin{figure}
 \centering
 \begin{tabular}{c |  c    }
\hline \hline    
    Scaling Factor             & $\alpha$   \\
    \hline 
n$^{(1)}$          & 1.7      \\
V$^{(1)}$ & 1.7      \\ 
H$^{(1)}$          & 1.6       \\ 
P$^{(1)}$          & 1.2    \\\hline 
Total              & 1.4          \\\hline \hline 
\end{tabular}
 \includegraphics[width=0.9\columnwidth]{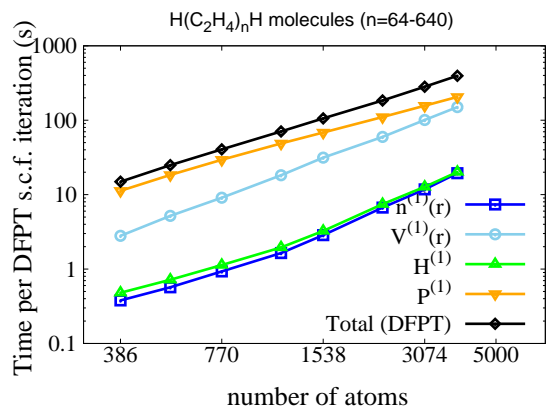}
 \caption{Dependence of the CPU time per DFPT cycle on the number of atoms in the H(C$_2$H$_4$)$_n$H molecules. The perturbation of the electric field is along the Z direction. Here, the total CPU time (black line) as well as its four components, \textit{i.e.}, CPU time for the density $n^{(1)}$ (blue line), the electrostatic potential $V^{(1)}$ (sky blue line), the Hamiltonian matrix $H^{(1)}$ (green line), and the density matrix $P^{(1)}$ (orange line), are shown. Double logarithmic axes are used. The fitted CPU time exponents~$\alpha$ for the H(C$_2$H$_4$)$_n$H molecules (n=64-640) are given in the table. The fits were performed using the expression~$t = c N^{\alpha}$ for the CPU time as function of the number of atoms~$N$. }
\label{fig:scaling_c2h4_mol}
 \end{figure}

The scalability tests are performed on the Tianhe-2 supercomputer
located at the National Supercomputing Center in Guangzhou, China. 
The largest number of nodes that we can use for performance test is 1,050 nodes~(25,200 cores). Each node is composed of two Intel Ivy Bridge E5-2692 processors (12 cores each at 2.2 GHz). 
%-------speedup----------
In Fig.~\ref{fig:speedup_mol}, we show the parallel scalability for 
the finite system containing 770 atoms. Here we can see the first order
density matrix calculation is the most time-consuming step, and the 
scalability is good.  A relative speedup of 15.2$\times$ is obtained reducing the wall-time per DFPT iteration from 81.5 sec on 24 MPI cores to around 5.3 sec on 768 MPI cores. 
And the parallel efficiency is nearly 47\% when using 768 cores. 
Beside the cluster systems under free boundary condition, we also investigate the parallel scalability for an extended system~(polyethylene) under periodic boundary conditions with a unit cell containing 768 atoms as shown in Fig. \ref{fig:speedup_pbc}. $\Gamma$-point is sufficient to sample the reciprocal space due to the large unit cell, in this system, the first order potential is the most time-consuming step, which shows almost ideal scaling, and the parallel efficiency is around 87\% when using 768 cores.

\begin{figure}
 \centering
 \includegraphics[width=0.9\columnwidth]{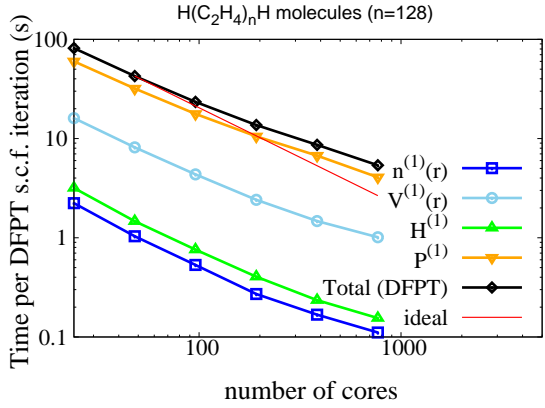}
 \caption{Parallel scalability for the H(C$_2$H$_4$)$_n$H molecule containing 770 atoms. The perturbation of the electric field is along the Z direction. Here, the total CPU time (black line) as well as its four components, \textit{i.e.}, CPU time for the density $n^{(1)}$ (blue line), the electrostatic potential $V^{(1)}$ (sky blue line), the Hamiltonian matrix $H^{(1)}$ (green line), and the density matrix $P^{(1)}$ (orange line), are shown. Double logarithmic axes are used. The red line corresponds to the ideal scaling. Here we use light settings for the integration, a ``tier 1'' basis set, and the LDA functional, NTPoly-filter~($10^{-6}$) and NTPoly-tolerance~($10^{-5}$) as the numerical thresholds.}
 \label{fig:speedup_mol}
 \end{figure} 
 
\begin{figure}
 \centering
 \includegraphics[width=0.9\columnwidth]{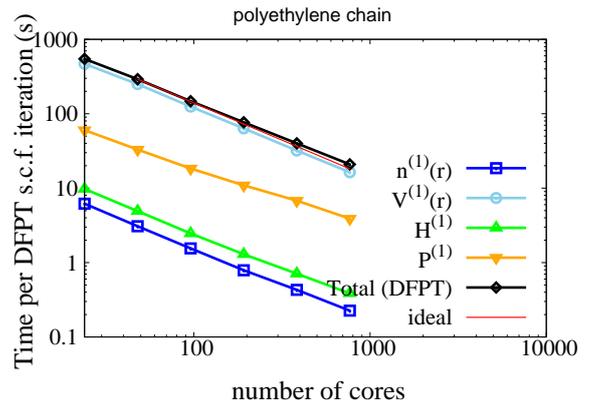}
 \caption{Parallel scalability for the extended system~(polyethylene) with a unit cell containing 768 atoms. 
The perturbation of the electric field is along the Z direction. Here, the total CPU time (black line) as well as its four components, \textit{i.e.}, CPU time for the density $n^{(1)}$ (blue line), the electrostatic potential $V^{(1)}$ (sky blue line), the Hamiltonian matrix $H^{(1)}$ (green line), and the density matrix $P^{(1)}$ (orange line), are shown. Double logarithmic axes are used. The red line corresponds to the ideal scaling. Here we use light settings for the integration, a ``tier 1'' basis set, and the LDA functional, NTPoly-filter~($10^{-6}$) and NTPoly-tolerance~($10^{-5}$) as the numerical thresholds.}
  \label{fig:speedup_pbc}
 \end{figure}

%-------stronge scaling-------
It should be noted that, the moderate parallel efficiency~(47\%) in Fig.~\ref{fig:speedup_mol} with 768 cores is because of low computational intensity with 770 atoms. If we increase the H(C$_2$H$_4$)$_n$H system size to 3074 atoms, then the parallel efficiency increases to around 55\% at 768 cores.    
In Fig.~\ref{fig:speedup}, we further investigate the parallel scalability of the TC2-CPSCF method with different system sizes for the H(C$_2$H$_4$)$_n$H molecules.  
Here we can see for system sizes ranged from 3,074 atoms to 12,290 atoms,  with the number of the basis functions changing from 24,586 to 98,314,  the scalability is good up to the maximum 25,200 CPU cores.   
For the system contained 3,074 atoms, a speedup of 22.3 $\times$ is obtained reducing the wall-time per DFPT iteration from 342.2 sec on 48 MPI cores to 15.3 sec on 3,072 MPI cores.
Then for a larger system contained 6,146 atoms, a speedup of 9.5 $\times$ is obtained reducing the wall-time per DFPT iteration from 253.9 sec on 240 MPI cores to 26.9 sec on 6,144 MPI cores. Finally, the total time per DFPT cycle speedups of up to 4.5$\times$ when we go from 2,400 cores to 25,200 cores for the H(C$_2$H$_4$)$_n$H system contained 12,290 atoms.

\begin{figure}
 \centering
 \includegraphics[width=0.9\columnwidth]{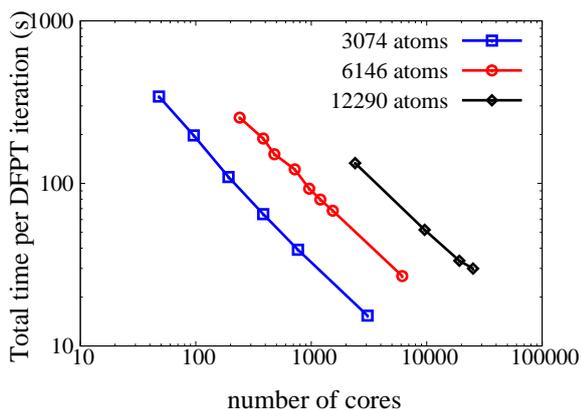}
 \caption{The strong scalability for the total time per DFPT iteration with different system sizes in the H(C$_2$H$_4$)$_n$H molecules. The systems we are using contained 3,074 atoms~(24,586 basis functions), 6,146 atoms~(49,162 basis functions) and 12,290 atoms~(98,314 basis functions).}
  \label{fig:speedup}
 \end{figure}

\section{Conclusions}
\label{sec:conclusions}
We have implemented an efficient parallel linear scaling method for
perturbations of homogeneous electric fields within an all-electron, numeric atom-centered orbitals framework.
We have validated the implementation by comparing polarizabilities of molecules calculated from this $O(N)$ approach with those obtained from the traditional $O(N^{3})$ method.
The results can be systematically converged with respect to the used numerical parameters.
The scaling exponent of the computation time in calculating the first order density matrix is $\alpha=1.2$ for system sizes up to thousands of atoms, which is close to the expected $O(N)$ scaling. The implemented TC2-CPSCF method exhibits a good parallel scalability that can be extended up to 25,200 cores in real systems. The formalism described in this paper could also be applied
in dealing with other type of perturbations, e.g. atomic displacements in the lattice dynamics. Moreover, 
the 3D-SpGEMM algorithm employed in this work can also be used in the density-matrix-based Laplace-transformed CPSCF method~\cite{Beer2008} and the density-matrix-based time-dependent self-consistent field method ~\cite{Kussmann2007} for calculating dynamic polarizabilities.

%%%%%%%%%%%%%%%%%%%%%%%%%%%%%%%%%%%%%%%%%%%%%%%%%%%%%%%%%%%%%%%%%%
%                        Bibliography                            %
%%%%%%%%%%%%%%%%%%%%%%%%%%%%%%%%%%%%%%%%%%%%%%%%%%%%%%%%%%%%%%%%%%
\bibliography{manuscript} % Produces the bibliography via BibTeX.

\begin{thebibliography}{55}
\expandafter\ifx\csname natexlab\endcsname\relax\def\natexlab#1{#1}\fi
\providecommand{\url}[1]{\texttt{#1}}
\providecommand{\href}[2]{#2}
\providecommand{\path}[1]{#1}
\providecommand{\DOIprefix}{doi:}
\providecommand{\ArXivprefix}{arXiv:}
\providecommand{\URLprefix}{URL: }
\providecommand{\Pubmedprefix}{pmid:}
\providecommand{\doi}[1]{\href{http://dx.doi.org/#1}{\path{#1}}}
\providecommand{\Pubmed}[1]{\href{pmid:#1}{\path{#1}}}
\providecommand{\bibinfo}[2]{#2}
\ifx\xfnm\relax \def\xfnm[#1]{\unskip,\space#1}\fi
%Type = Article
\bibitem[{Hohenberg(1964)}]{Hohenberg1964}
\bibinfo{author}{P.~Hohenberg}, \bibinfo{journal}{Phys. Rev.}
  \bibinfo{volume}{136} (\bibinfo{year}{1964}) \bibinfo{pages}{B864--B871}.
  \URLprefix \url{http://link.aps.org/doi/10.1103/PhysRev.136.B864}.
  \DOIprefix\doi{10.1103/PhysRev.136.B864}.
%Type = Article
\bibitem[{Kohn and Sham(1965)}]{Kohn1965}
\bibinfo{author}{W.~Kohn}, \bibinfo{author}{L.~J. Sham},
  \bibinfo{journal}{Phys. Rev.} \bibinfo{volume}{140} (\bibinfo{year}{1965})
  \bibinfo{pages}{A1133--A1138}. \URLprefix
  \url{http://link.aps.org/doi/10.1103/PhysRev.140.A1133}.
  \DOIprefix\doi{10.1103/PhysRev.140.A1133}.
%Type = Article
\bibitem[{Gonze(1997)}]{Gonze1997-1}
\bibinfo{author}{X.~Gonze}, \bibinfo{journal}{Phys. Rev. B}
  \bibinfo{volume}{55} (\bibinfo{year}{1997}) \bibinfo{pages}{10337--10354}.
  \URLprefix \url{http://link.aps.org/doi/10.1103/PhysRevB.55.10337}.
  \DOIprefix\doi{10.1103/PhysRevB.55.10337}.
%Type = Article
\bibitem[{Gonze and Lee(1997)}]{Gonze1997-2}
\bibinfo{author}{X.~Gonze}, \bibinfo{author}{C.~Lee}, \bibinfo{journal}{Phys.
  Rev. B} \bibinfo{volume}{55} (\bibinfo{year}{1997})
  \bibinfo{pages}{10355--10368}. \URLprefix
  \url{http://link.aps.org/doi/10.1103/PhysRevB.55.10355}.
  \DOIprefix\doi{10.1103/PhysRevB.55.10355}.
%Type = Article
\bibitem[{Baroni et~al.(2001)Baroni, de~Gironcoli, Dal~Corso, and
  Giannozzi}]{Baroni-2001}
\bibinfo{author}{S.~Baroni}, \bibinfo{author}{S.~de~Gironcoli},
  \bibinfo{author}{A.~Dal~Corso}, \bibinfo{author}{P.~Giannozzi},
  \bibinfo{journal}{Rev. Mod. Phys.} \bibinfo{volume}{73}
  (\bibinfo{year}{2001}) \bibinfo{pages}{515--562}. \URLprefix
  \url{http://link.aps.org/doi/10.1103/RevModPhys.73.515}.
  \DOIprefix\doi{10.1103/RevModPhys.73.515}.
%Type = Article
\bibitem[{Gerratt and Mills(1968)}]{Gerratt-1967}
\bibinfo{author}{J.~Gerratt}, \bibinfo{author}{I.~M. Mills},
  \bibinfo{journal}{J. Chem. Phys.} \bibinfo{volume}{49} (\bibinfo{year}{1968})
  \bibinfo{pages}{1719--1729}. \URLprefix
  \url{http://link.aip.org/link/?JCP/49/1719/1}.
  \DOIprefix\doi{10.1063/1.1670299}.
%Type = Article
\bibitem[{Pople et~al.(1979)Pople, Krishnan, Schlegel, and
  Binkley}]{Pople-1979}
\bibinfo{author}{J.~A. Pople}, \bibinfo{author}{R.~Krishnan},
  \bibinfo{author}{H.~B. Schlegel}, \bibinfo{author}{J.~S. Binkley},
  \bibinfo{journal}{International Journal of Quantum Chemistry}
  \bibinfo{volume}{16} (\bibinfo{year}{1979}) \bibinfo{pages}{225--241}.
  \URLprefix \url{http://dx.doi.org/10.1002/qua.560160825}.
  \DOIprefix\doi{10.1002/qua.560160825}.
%Type = Article
\bibitem[{Dykstra and Jasien(1984)}]{Dykstra-1984}
\bibinfo{author}{C.~E. Dykstra}, \bibinfo{author}{P.~G. Jasien},
  \bibinfo{journal}{Chem. Phys. Lett.} \bibinfo{volume}{109}
  (\bibinfo{year}{1984}) \bibinfo{pages}{388 -- 393}. \URLprefix
  \url{http://www.sciencedirect.com/science/article/pii/0009261484856079}.
  \DOIprefix\doi{10.1016/0009-2614(84)85607-9}.
%Type = Article
\bibitem[{Frisch et~al.(1990)Frisch, Head-Gordon, and Pople}]{Frisch-1990}
\bibinfo{author}{M.~Frisch}, \bibinfo{author}{M.~Head-Gordon},
  \bibinfo{author}{J.~Pople}, \bibinfo{journal}{Chem. Phys.}
  \bibinfo{volume}{141} (\bibinfo{year}{1990}) \bibinfo{pages}{189 -- 196}.
  \URLprefix
  \url{http://www.sciencedirect.com/science/article/pii/030101049087055G}.
  \DOIprefix\doi{10.1016/0301-0104(90)87055-G}.
%Type = Article
\bibitem[{Ochsenfeld and Head-Gordon(1997)}]{Ochsenfeld-1997}
\bibinfo{author}{C.~Ochsenfeld}, \bibinfo{author}{M.~Head-Gordon},
  \bibinfo{journal}{Chem. Phys. Lett.} \bibinfo{volume}{270}
  (\bibinfo{year}{1997}) \bibinfo{pages}{399 -- 405}. \URLprefix
  \url{http://www.sciencedirect.com/science/article/pii/S0009261497004028}.
  \DOIprefix\doi{10.1016/S0009-2614(97)00402-8}.
%Type = Article
\bibitem[{Liang et~al.(2005)Liang, Zhao, and Head-Gordon}]{Liang-2005}
\bibinfo{author}{W.~Liang}, \bibinfo{author}{Y.~Zhao},
  \bibinfo{author}{M.~Head-Gordon}, \bibinfo{journal}{J. Chem. Phys.}
  \bibinfo{volume}{123} (\bibinfo{year}{2005}) \bibinfo{pages}{194106}.
  \URLprefix \url{http://link.aip.org/link/?JCP/123/194106/1}.
  \DOIprefix\doi{10.1063/1.2114847}.
%Type = Article
\bibitem[{Shang et~al.(2017)Shang, Carbogno, Rinke, and Scheffler}]{Shang2017}
\bibinfo{author}{H.~Shang}, \bibinfo{author}{C.~Carbogno},
  \bibinfo{author}{P.~Rinke}, \bibinfo{author}{M.~Scheffler},
  \bibinfo{journal}{Computer Physics Communications} \bibinfo{volume}{215}
  (\bibinfo{year}{2017}) \bibinfo{pages}{26--46}. \URLprefix
  \url{http://www.sciencedirect.com/science/article/pii/S0010465517300437
  https://linkinghub.elsevier.com/retrieve/pii/S0010465517300437}.
  \DOIprefix\doi{10.1016/j.cpc.2017.02.001}.
%Type = Article
\bibitem[{Shang et~al.(2018)Shang, Raimbault, Rinke, Scheffler, Rossi, and
  Carbogno}]{Shang2018}
\bibinfo{author}{H.~Shang}, \bibinfo{author}{N.~Raimbault},
  \bibinfo{author}{P.~Rinke}, \bibinfo{author}{M.~Scheffler},
  \bibinfo{author}{M.~Rossi}, \bibinfo{author}{C.~Carbogno},
  \bibinfo{journal}{New Journal of Physics} \bibinfo{volume}{20}
  (\bibinfo{year}{2018}) \bibinfo{pages}{073040}. \URLprefix
  \url{http://stacks.iop.org/1367-2630/20/i=7/a=073040?key=crossref.b45b8680fc0308226fe0611417a68450}.
  \DOIprefix\doi{10.1088/1367-2630/aace6d}.
%Type = Article
\bibitem[{Bowler and Miyazaki(2011)}]{Bowler2011}
\bibinfo{author}{D.~R. Bowler}, \bibinfo{author}{T.~Miyazaki},
  \bibinfo{journal}{Reports on Progress in Physics} \bibinfo{volume}{75}
  (\bibinfo{year}{2011}) \bibinfo{pages}{036503}. \URLprefix
  \url{http://stacks.iop.org/0034-4885/75/i=3/a=036503?key=crossref.28e3cb0ce7d7274e9d63f9158aff7224
  http://arxiv.org/abs/1108.5976
  http://dx.doi.org/10.1088/0034-4885/75/3/036503}.
  \DOIprefix\doi{10.1088/0034-4885/75/3/036503}.
  \href{http://arxiv.org/abs/1108.5976}{\tt arXiv:1108.5976}.
%Type = Article
\bibitem[{Kohn(1996)}]{Kohn1996}
\bibinfo{author}{W.~Kohn}, \bibinfo{journal}{Phys. Rev. Lett.}
  \bibinfo{volume}{76} (\bibinfo{year}{1996}) \bibinfo{pages}{3168--3171}.
  \URLprefix \url{https://link.aps.org/doi/10.1103/PhysRevLett.76.3168}.
  \DOIprefix\doi{10.1103/PhysRevLett.76.3168}.
%Type = Article
\bibitem[{Soler et~al.(2002)Soler, Artacho, Gale, Garc\'{\i}a, Junquera,
  Ordej\'{o}n, and S\'{a}nchez-Portal}]{Soler2002}
\bibinfo{author}{J.~M. Soler}, \bibinfo{author}{E.~Artacho},
  \bibinfo{author}{J.~D. Gale}, \bibinfo{author}{A.~Garc\'{\i}a},
  \bibinfo{author}{J.~Junquera}, \bibinfo{author}{P.~Ordej\'{o}n},
  \bibinfo{author}{D.~S\'{a}nchez-Portal}, \bibinfo{journal}{J. Phys. Condens.
  Matter} \bibinfo{volume}{14} (\bibinfo{year}{2002})
  \bibinfo{pages}{2745--2779}. \URLprefix
  \url{http://iopscience.iop.org/0953-8984/14/11/302/}.
  \DOIprefix\doi{10.1088/0953-8984/14/11/302}.
%Type = Article
\bibitem[{Bowler and Miyazaki(2010)}]{Bowler2010}
\bibinfo{author}{D.~R. Bowler}, \bibinfo{author}{T.~Miyazaki},
  \bibinfo{journal}{J. Phys. Condens. Matter} \bibinfo{volume}{22}
  (\bibinfo{year}{2010}) \bibinfo{pages}{74207}. \URLprefix
  \url{http://stacks.iop.org/0953-8984/22/i=7/a=074207}.
%Type = Article
\bibitem[{Shang et~al.(2010)Shang, Xiang, Li, and Yang}]{Shang2010a}
\bibinfo{author}{H.~Shang}, \bibinfo{author}{H.~Xiang},
  \bibinfo{author}{Z.~Li}, \bibinfo{author}{J.~Yang},
  \bibinfo{journal}{International Reviews in Physical Chemistry}
  \bibinfo{volume}{29} (\bibinfo{year}{2010}) \bibinfo{pages}{665--691}.
  \URLprefix
  \url{http://www.tandfonline.com/doi/abs/10.1080/0144235X.2010.520454}.
  \DOIprefix\doi{10.1080/0144235X.2010.520454}.
%Type = Article
\bibitem[{Torralba et~al.(2008)Torralba, Todorovi{\'{c}}, Br{\'{a}}zdov{\'{a}},
  Choudhury, Miyazaki, Gillan, and Bowler}]{CONQUEST}
\bibinfo{author}{A.~S. Torralba}, \bibinfo{author}{M.~Todorovi{\'{c}}},
  \bibinfo{author}{V.~Br{\'{a}}zdov{\'{a}}}, \bibinfo{author}{R.~Choudhury},
  \bibinfo{author}{T.~Miyazaki}, \bibinfo{author}{M.~J. Gillan},
  \bibinfo{author}{D.~R. Bowler}, \bibinfo{journal}{Journal of Physics:
  Condensed Matter} \bibinfo{volume}{20} (\bibinfo{year}{2008})
  \bibinfo{pages}{294206}. \URLprefix
  \url{https://doi.org/10.1088\%2F0953-8984\%2F20\%2F29\%2F294206}.
  \DOIprefix\doi{10.1088/0953-8984/20/29/294206}.
%Type = Article
\bibitem[{Weber and Challacombe(2006)}]{Weber2006}
\bibinfo{author}{V.~Weber}, \bibinfo{author}{M.~Challacombe},
  \bibinfo{journal}{The Journal of Chemical Physics} \bibinfo{volume}{125}
  (\bibinfo{year}{2006}) \bibinfo{pages}{104110}. \URLprefix
  \url{https://doi.org/10.1063/1.2222359}. \DOIprefix\doi{10.1063/1.2222359}.
  \href{http://arxiv.org/abs/https://doi.org/10.1063/1.2222359}{\tt
  arXiv:https://doi.org/10.1063/1.2222359}.
%Type = Article
\bibitem[{Wu et~al.(2009)Wu, Selloni, and Car}]{Wu2009}
\bibinfo{author}{X.~Wu}, \bibinfo{author}{A.~Selloni},
  \bibinfo{author}{R.~Car}, \bibinfo{journal}{Phys. Rev. B}
  \bibinfo{volume}{79} (\bibinfo{year}{2009}) \bibinfo{pages}{085102}.
  \URLprefix \url{https://link.aps.org/doi/10.1103/PhysRevB.79.085102}.
  \DOIprefix\doi{10.1103/PhysRevB.79.085102}.
%Type = Article
\bibitem[{Niklasson(2002)}]{Niklasson2002PRB}
\bibinfo{author}{A.~M.~N. Niklasson}, \bibinfo{journal}{Phys. Rev. B}
  \bibinfo{volume}{66} (\bibinfo{year}{2002}) \bibinfo{pages}{155115}.
  \URLprefix
  \url{https://link-aps-org-443.webvpn.las.ac.cn/doi/10.1103/PhysRevB.66.155115}.
  \DOIprefix\doi{10.1103/PhysRevB.66.155115}.
%Type = Article
\bibitem[{Niklasson and Challacombe(2004)}]{Niklasson2004}
\bibinfo{author}{A.~M.~N. Niklasson}, \bibinfo{author}{M.~Challacombe},
  \bibinfo{journal}{Physical Review Letters} \bibinfo{volume}{92}
  (\bibinfo{year}{2004}) \bibinfo{pages}{193001}. \URLprefix
  \url{https://link.aps.org/doi/10.1103/PhysRevLett.92.193001}.
  \DOIprefix\doi{10.1103/PhysRevLett.92.193001}.
  \href{http://arxiv.org/abs/0311591}{\tt arXiv:0311591}.
%Type = Article
\bibitem[{Mniszewski et~al.(2015)Mniszewski, Cawkwell, Wall, Mohd-Yusof, Bock,
  Germann, and Niklasson}]{Mniszewski2015}
\bibinfo{author}{S.~M. Mniszewski}, \bibinfo{author}{M.~J. Cawkwell},
  \bibinfo{author}{M.~E. Wall}, \bibinfo{author}{J.~Mohd-Yusof},
  \bibinfo{author}{N.~Bock}, \bibinfo{author}{T.~C. Germann},
  \bibinfo{author}{A.~M.~N. Niklasson}, \bibinfo{journal}{Journal of Chemical
  Theory and Computation} \bibinfo{volume}{11} (\bibinfo{year}{2015})
  \bibinfo{pages}{4644--4654}. \URLprefix
  \url{https://pubs.acs.org/doi/10.1021/acs.jctc.5b00552}.
  \DOIprefix\doi{10.1021/acs.jctc.5b00552}.
%Type = Article
\bibitem[{Cawkwell et~al.(2014)Cawkwell, Wood, Niklasson, and
  Mniszewski}]{Cawkwell2014}
\bibinfo{author}{M.~J. Cawkwell}, \bibinfo{author}{M.~A. Wood},
  \bibinfo{author}{A.~M. Niklasson}, \bibinfo{author}{S.~M. Mniszewski},
  \bibinfo{journal}{Journal of Chemical Theory and Computation}
  \bibinfo{volume}{10} (\bibinfo{year}{2014}) \bibinfo{pages}{5391--5396}.
  \DOIprefix\doi{10.1021/ct5008229}.
%Type = Inproceedings
\bibitem[{Lazzaro et~al.(2017)Lazzaro, VandeVondele, Hutter, and
  Sch\"{u}tt}]{Lazzaro2017}
\bibinfo{author}{A.~Lazzaro}, \bibinfo{author}{J.~VandeVondele},
  \bibinfo{author}{J.~Hutter}, \bibinfo{author}{O.~Sch\"{u}tt}, in:
  \bibinfo{booktitle}{Proceedings of the Platform for Advanced Scientific
  Computing Conference}, PASC ’17, \bibinfo{publisher}{Association for
  Computing Machinery}, \bibinfo{address}{New York, NY, USA},
  \bibinfo{year}{2017}. \URLprefix
  \url{https://doi.org/10.1145/3093172.3093228}.
  \DOIprefix\doi{10.1145/3093172.3093228}.
%Type = Article
\bibitem[{Hutter et~al.(2014)Hutter, Iannuzzi, Schiffmann, and
  VandeVondele}]{Hutter2014}
\bibinfo{author}{J.~Hutter}, \bibinfo{author}{M.~Iannuzzi},
  \bibinfo{author}{F.~Schiffmann}, \bibinfo{author}{J.~VandeVondele},
  \bibinfo{journal}{WIREs Computational Molecular Science} \bibinfo{volume}{4}
  (\bibinfo{year}{2014}) \bibinfo{pages}{15--25}. \URLprefix
  \url{https://www.onlinelibrary.wiley.com/doi/abs/10.1002/wcms.1159}.
  \DOIprefix\doi{10.1002/wcms.1159}.
  \href{http://arxiv.org/abs/https://www.onlinelibrary.wiley.com/doi/pdf/10.1002/wcms.1159}{\tt
  arXiv:https://www.onlinelibrary.wiley.com/doi/pdf/10.1002/wcms.1159}.
%Type = Article
\bibitem[{Azad et~al.(2016)Azad, Ballard, Bulu{\c{c}}, Demmel, Grigori,
  Schwartz, Toledo, and Williams}]{Azad2016}
\bibinfo{author}{A.~Azad}, \bibinfo{author}{G.~Ballard},
  \bibinfo{author}{A.~Bulu{\c{c}}}, \bibinfo{author}{J.~Demmel},
  \bibinfo{author}{L.~Grigori}, \bibinfo{author}{O.~Schwartz},
  \bibinfo{author}{S.~Toledo}, \bibinfo{author}{S.~Williams},
  \bibinfo{journal}{SIAM Journal on Scientific Computing} \bibinfo{volume}{38}
  (\bibinfo{year}{2016}) \bibinfo{pages}{C624--C651}. \URLprefix
  \url{http://www.siam.org/journals/sisc/38-6/M104253.html
  http://epubs.siam.org/doi/10.1137/15M104253X}.
  \DOIprefix\doi{10.1137/15M104253X}.
%Type = Article
\bibitem[{Dawson and Nakajima(2018)}]{Dawson2018}
\bibinfo{author}{W.~Dawson}, \bibinfo{author}{T.~Nakajima},
  \bibinfo{journal}{Computer Physics Communications} \bibinfo{volume}{225}
  (\bibinfo{year}{2018}) \bibinfo{pages}{154--165}. \URLprefix
  \url{https://doi.org/10.1016/j.cpc.2017.12.010
  https://linkinghub.elsevier.com/retrieve/pii/S0010465517304150}.
  \DOIprefix\doi{10.1016/j.cpc.2017.12.010}.
%Type = Article
\bibitem[{Weber et~al.(2004)Weber, Niklasson, and Challacombe}]{Weber2004}
\bibinfo{author}{V.~Weber}, \bibinfo{author}{A.~M.~N. Niklasson},
  \bibinfo{author}{M.~Challacombe}, \bibinfo{journal}{Physical Review Letters}
  \bibinfo{volume}{92} (\bibinfo{year}{2004}) \bibinfo{pages}{193002}.
  \URLprefix \url{https://link.aps.org/doi/10.1103/PhysRevLett.92.193002}.
  \DOIprefix\doi{10.1103/PhysRevLett.92.193002}.
  \href{http://arxiv.org/abs/0312634}{\tt arXiv:0312634}.
%Type = Article
\bibitem[{Xiang et~al.(2006)Xiang, Yang, Hou, and Zhu}]{Xiang2006}
\bibinfo{author}{H.~J. Xiang}, \bibinfo{author}{J.~Yang},
  \bibinfo{author}{J.~G. Hou}, \bibinfo{author}{Q.~Zhu},
  \bibinfo{journal}{Physical Review Letters} \bibinfo{volume}{97}
  (\bibinfo{year}{2006}) \bibinfo{pages}{266402}. \URLprefix
  \url{http://link.aps.org/doi/10.1103/PhysRevLett.97.266402
  https://link.aps.org/doi/10.1103/PhysRevLett.97.266402}.
  \DOIprefix\doi{10.1103/PhysRevLett.97.266402}.
%Type = Article
\bibitem[{Blum et~al.(2009)Blum, Gehrke, Hanke, Havu, Havu, Ren, Reuter, and
  Scheffler}]{Blum2009}
\bibinfo{author}{V.~Blum}, \bibinfo{author}{R.~Gehrke},
  \bibinfo{author}{F.~Hanke}, \bibinfo{author}{P.~Havu},
  \bibinfo{author}{V.~Havu}, \bibinfo{author}{X.~Ren},
  \bibinfo{author}{K.~Reuter}, \bibinfo{author}{M.~Scheffler},
  \bibinfo{journal}{Comput. Phys. Commun.} \bibinfo{volume}{180}
  (\bibinfo{year}{2009}) \bibinfo{pages}{2175--2196}. \URLprefix
  \url{http://linkinghub.elsevier.com/retrieve/pii/S0010465509002033}.
  \DOIprefix\doi{10.1016/j.cpc.2009.06.022}.
%Type = Article
\bibitem[{Delley(1991)}]{Delley1991}
\bibinfo{author}{B.~Delley}, \bibinfo{journal}{J. Chem. Phys.}
  \bibinfo{volume}{94} (\bibinfo{year}{1991}) \bibinfo{pages}{7245}. \URLprefix
  \url{http://scitation.aip.org/content/aip/journal/jcp/94/11/10.1063/1.460208}.
  \DOIprefix\doi{10.1063/1.460208}.
%Type = Article
\bibitem[{Delley(1990)}]{Delley-partition}
\bibinfo{author}{B.~Delley}, \bibinfo{journal}{J. Chem. Phys.}
  \bibinfo{volume}{92} (\bibinfo{year}{1990}) \bibinfo{pages}{508}. \URLprefix
  \url{http://scitation.aip.org/content/aip/journal/jcp/92/1/10.1063/1.458452}.
  \DOIprefix\doi{10.1063/1.458452}.
%Type = Article
\bibitem[{Gonze and Vigneron(1989)}]{Gonze-1989}
\bibinfo{author}{X.~Gonze}, \bibinfo{author}{J.-P. Vigneron},
  \bibinfo{journal}{Phys. Rev. B} \bibinfo{volume}{39} (\bibinfo{year}{1989})
  \bibinfo{pages}{13120--13128}. \URLprefix
  \url{http://link.aps.org/doi/10.1103/PhysRevB.39.13120}.
  \DOIprefix\doi{10.1103/PhysRevB.39.13120}.
%Type = Article
\bibitem[{Sternheimer(1954)}]{Sternheimer1954}
\bibinfo{author}{R.~M. Sternheimer}, \bibinfo{journal}{Phys. Rev.}
  \bibinfo{volume}{96} (\bibinfo{year}{1954}) \bibinfo{pages}{951--968}.
  \URLprefix \url{http://link.aps.org/doi/10.1103/PhysRev.96.951}.
  \DOIprefix\doi{10.1103/PhysRev.96.951}.
%Type = Article
\bibitem[{Niklasson et~al.(2003)Niklasson, Tymczak, and
  Challacombe}]{Niklasson2003JCP}
\bibinfo{author}{A.~M.~N. Niklasson}, \bibinfo{author}{C.~J. Tymczak},
  \bibinfo{author}{M.~Challacombe}, \bibinfo{journal}{The Journal of Chemical
  Physics} \bibinfo{volume}{118} (\bibinfo{year}{2003})
  \bibinfo{pages}{8611--8620}. \URLprefix
  \url{https://doi-org-443.webvpn.las.ac.cn/10.1063/1.1559913}.
  \DOIprefix\doi{10.1063/1.1559913}.
  \href{http://arxiv.org/abs/https://doi-org-443.webvpn.las.ac.cn/10.1063/1.1559913}{\tt
  arXiv:https://doi-org-443.webvpn.las.ac.cn/10.1063/1.1559913}.
%Type = Article
\bibitem[{Niklasson and Weber(2007)}]{Niklasson2007}
\bibinfo{author}{A.~M.~N. Niklasson}, \bibinfo{author}{V.~Weber},
  \bibinfo{journal}{The Journal of Chemical Physics} \bibinfo{volume}{127}
  (\bibinfo{year}{2007}) \bibinfo{pages}{064105}. \URLprefix
  \url{http://aip.scitation.org/doi/10.1063/1.2755775}.
  \DOIprefix\doi{10.1063/1.2755775}.
%Type = Article
\bibitem[{Rudberg and Rubensson(2011)}]{Rudberg2011}
\bibinfo{author}{E.~Rudberg}, \bibinfo{author}{E.~H. Rubensson},
  \bibinfo{journal}{Journal of Physics: Condensed Matter} \bibinfo{volume}{23}
  (\bibinfo{year}{2011}) \bibinfo{pages}{075502}. \URLprefix
  \url{http://stacks.iop.org/0953-8984/23/i=7/a=075502?key=crossref.f46001d5e55d398d8173874c06aff985}.
  \DOIprefix\doi{10.1088/0953-8984/23/7/075502}.
%Type = Article
\bibitem[{Bock et~al.(2018)Bock, Negre, Mniszewski, Mohd-Yusof, Aradi,
  Fattebert, Osei-Kuffuor, Germann, and Niklasson}]{Bock2018}
\bibinfo{author}{N.~Bock}, \bibinfo{author}{C.~F.~A. Negre},
  \bibinfo{author}{S.~M. Mniszewski}, \bibinfo{author}{J.~Mohd-Yusof},
  \bibinfo{author}{B.~Aradi}, \bibinfo{author}{J.-L. Fattebert},
  \bibinfo{author}{D.~Osei-Kuffuor}, \bibinfo{author}{T.~C. Germann},
  \bibinfo{author}{A.~M.~N. Niklasson}, \bibinfo{journal}{The Journal of
  Supercomputing} \bibinfo{volume}{74} (\bibinfo{year}{2018})
  \bibinfo{pages}{6201--6219}. \URLprefix
  \url{http://link.springer.com/10.1007/s11227-018-2533-0}.
  \DOIprefix\doi{10.1007/s11227-018-2533-0}.
%Type = Article
\bibitem[{L\"{o}wdin(1950)}]{Lowdin1950}
\bibinfo{author}{P.~L\"{o}wdin}, \bibinfo{journal}{The Journal of Chemical
  Physics} \bibinfo{volume}{18} (\bibinfo{year}{1950})
  \bibinfo{pages}{365--375}. \URLprefix
  \url{https://doi-org-443.webvpn.las.ac.cn/10.1063/1.1747632}.
  \DOIprefix\doi{10.1063/1.1747632}.
  \href{http://arxiv.org/abs/https://doi-org-443.webvpn.las.ac.cn/10.1063/1.1747632}{\tt
  arXiv:https://doi-org-443.webvpn.las.ac.cn/10.1063/1.1747632}.
%Type = Article
\bibitem[{L\"{o}wdin(1956)}]{Lowdin1956}
\bibinfo{author}{P.-O. L\"{o}wdin}, \bibinfo{journal}{Advances in Physics}
  \bibinfo{volume}{5} (\bibinfo{year}{1956}) \bibinfo{pages}{1--171}.
  \URLprefix \url{https://doi.org/10.1080/00018735600101155}.
  \DOIprefix\doi{10.1080/00018735600101155}.
  \href{http://arxiv.org/abs/https://doi.org/10.1080/00018735600101155}{\tt
  arXiv:https://doi.org/10.1080/00018735600101155}.
%Type = Article
\bibitem[{Jans{\'{i}}k et~al.(2007)Jans{\'{i}}k, H{\o}st, J{\o}rgensen, Olsen,
  and Helgaker}]{Jansik2007}
\bibinfo{author}{B.~Jans{\'{i}}k}, \bibinfo{author}{S.~H{\o}st},
  \bibinfo{author}{P.~J{\o}rgensen}, \bibinfo{author}{J.~Olsen},
  \bibinfo{author}{T.~Helgaker}, \bibinfo{journal}{The Journal of Chemical
  Physics} \bibinfo{volume}{126} (\bibinfo{year}{2007})
  \bibinfo{pages}{124104}. \URLprefix \url{https://doi.org/10.1063/1.2709881}.
  \DOIprefix\doi{10.1063/1.2709881}.
%Type = Article
\bibitem[{Ren et~al.(2012)Ren, Rinke, Blum, Wieferink, Tkatchenko, Sanfilippo,
  Reuter, and Scheffler}]{Ren/etal:2012}
\bibinfo{author}{X.~Ren}, \bibinfo{author}{P.~Rinke},
  \bibinfo{author}{V.~Blum}, \bibinfo{author}{J.~Wieferink},
  \bibinfo{author}{A.~Tkatchenko}, \bibinfo{author}{A.~Sanfilippo},
  \bibinfo{author}{K.~Reuter}, \bibinfo{author}{M.~Scheffler},
  \bibinfo{journal}{New J. Phys.} \bibinfo{volume}{14} (\bibinfo{year}{2012})
  \bibinfo{pages}{053020}. \URLprefix
  \url{http://stacks.iop.org/1367-2630/14/i=5/a=053020?key=crossref.351b343783c2c1df1596219a941a74eb}.
  \DOIprefix\doi{10.1088/1367-2630/14/5/053020}.
%Type = Article
\bibitem[{Havu et~al.(2009)Havu, Blum, Havu, and Scheffler}]{Havu/etal:2009}
\bibinfo{author}{V.~Havu}, \bibinfo{author}{V.~Blum},
  \bibinfo{author}{P.~Havu}, \bibinfo{author}{M.~Scheffler},
  \bibinfo{journal}{J. Comput. Phys.} \bibinfo{volume}{228}
  (\bibinfo{year}{2009}) \bibinfo{pages}{8367--8379}. \URLprefix
  \url{http://linkinghub.elsevier.com/retrieve/pii/S0021999109004458}.
  \DOIprefix\doi{10.1016/j.jcp.2009.08.008}.
%Type = Article
\bibitem[{Gustavson(1978)}]{Gustavson1978}
\bibinfo{author}{F.~G. Gustavson}, \bibinfo{journal}{ACM Transactions on
  Mathematical Software (TOMS)} \bibinfo{volume}{4} (\bibinfo{year}{1978})
  \bibinfo{pages}{250--269}. \URLprefix
  \url{http://dl.acm.org/doi/10.1145/355791.355796}.
  \DOIprefix\doi{10.1145/355791.355796}.
%Type = Inproceedings
\bibitem[{Ballard et~al.(2013)Ballard, Buluc, Demmel, Grigori, Lipshitz,
  Schwartz, and Toledo}]{Ballard2013}
\bibinfo{author}{G.~Ballard}, \bibinfo{author}{A.~Buluc},
  \bibinfo{author}{J.~Demmel}, \bibinfo{author}{L.~Grigori},
  \bibinfo{author}{B.~Lipshitz}, \bibinfo{author}{O.~Schwartz},
  \bibinfo{author}{S.~Toledo}, in: \bibinfo{booktitle}{Proceedings of the 25th
  ACM symposium on Parallelism in algorithms and architectures - SPAA '13},
  \bibinfo{number}{2}, \bibinfo{publisher}{ACM Press}, \bibinfo{address}{New
  York, New York, USA}, \bibinfo{year}{2013}, p. \bibinfo{pages}{222}.
  \URLprefix \url{http://dl.acm.org/citation.cfm?doid=2486159.2486196}.
  \DOIprefix\doi{10.1145/2486159.2486196}.
%Type = Article
\bibitem[{Gilbert;(2011)}]{Bulu2011}
\bibinfo{author}{A.~B.~R. Gilbert;}, \bibinfo{journal}{The International
  Journal of High Performance Computing Applications} \bibinfo{volume}{25}
  (\bibinfo{year}{2011}) \bibinfo{pages}{496--509}. \URLprefix
  \url{http://dx.doi.org/10.1177/1094342011403516}.
  \DOIprefix\doi{10.1177/1094342011403516}.
  \href{http://arxiv.org/abs/http://dx.doi.org/10.1177/1094342011403516}{\tt
  arXiv:http://dx.doi.org/10.1177/1094342011403516}.
%Type = Article
\bibitem[{zhe Yu et~al.(2018)zhe Yu, Corsetti, García, Huhn, Jacquelin, Jia,
  Lange, Lin, Lu, Mi, Seifitokaldani, Álvaro Vázquez-Mayagoitia, Yang, Yang,
  and Blum}]{YU2018}
\bibinfo{author}{V.~W. zhe Yu}, \bibinfo{author}{F.~Corsetti},
  \bibinfo{author}{A.~García}, \bibinfo{author}{W.~P. Huhn},
  \bibinfo{author}{M.~Jacquelin}, \bibinfo{author}{W.~Jia},
  \bibinfo{author}{B.~Lange}, \bibinfo{author}{L.~Lin},
  \bibinfo{author}{J.~Lu}, \bibinfo{author}{W.~Mi},
  \bibinfo{author}{A.~Seifitokaldani}, \bibinfo{author}{Álvaro
  Vázquez-Mayagoitia}, \bibinfo{author}{C.~Yang}, \bibinfo{author}{H.~Yang},
  \bibinfo{author}{V.~Blum}, \bibinfo{journal}{Computer Physics Communications}
  \bibinfo{volume}{222} (\bibinfo{year}{2018}) \bibinfo{pages}{267 -- 285}.
  \URLprefix
  \url{http://www.sciencedirect.com/science/article/pii/S0010465517302941}.
  \DOIprefix\doi{https://doi.org/10.1016/j.cpc.2017.09.007}.
%Type = Article
\bibitem[{Rubensson and Sa{\l}ek(2005)}]{Rubensson2005}
\bibinfo{author}{E.~H. Rubensson}, \bibinfo{author}{P.~Sa{\l}ek},
  \bibinfo{journal}{Journal of Computational Chemistry} \bibinfo{volume}{26}
  (\bibinfo{year}{2005}) \bibinfo{pages}{1628--1637}. \URLprefix
  \url{http://doi.wiley.com/10.1002/jcc.20315}.
  \DOIprefix\doi{10.1002/jcc.20315}.
%Type = Article
\bibitem[{Rubensson et~al.(2008)Rubensson, Rudberg, and
  Sa{\l}ek}]{Rubensson2008}
\bibinfo{author}{E.~H. Rubensson}, \bibinfo{author}{E.~Rudberg},
  \bibinfo{author}{P.~Sa{\l}ek}, \bibinfo{journal}{The Journal of Chemical
  Physics} \bibinfo{volume}{128} (\bibinfo{year}{2008})
  \bibinfo{pages}{074106}. \URLprefix
  \url{http://aip.scitation.org/doi/10.1063/1.2826343}.
  \DOIprefix\doi{10.1063/1.2826343}.
%Type = Article
\bibitem[{Perdew and Zunger(1981)}]{Perdew/Zunger:1981}
\bibinfo{author}{J.~P. Perdew}, \bibinfo{author}{A.~Zunger},
  \bibinfo{journal}{Phys. Rev. B} \bibinfo{volume}{23} (\bibinfo{year}{1981})
  \bibinfo{pages}{5048--5079}. \URLprefix
  \url{http://link.aps.org/doi/10.1103/PhysRevB.23.5048}.
  \DOIprefix\doi{10.1103/PhysRevB.23.5048}.
%Type = Article
\bibitem[{Ceperley and Alder(1980)}]{Ceperley/Alder:1980}
\bibinfo{author}{D.~M. Ceperley}, \bibinfo{author}{B.~J. Alder},
  \bibinfo{journal}{Phys. Rev. Lett.} \bibinfo{volume}{45}
  (\bibinfo{year}{1980}) \bibinfo{pages}{566--569}. \URLprefix
  \url{http://link.aps.org/doi/10.1103/PhysRevLett.45.566}.
  \DOIprefix\doi{10.1103/PhysRevLett.45.566}.
%Type = Article
\bibitem[{Beer and Ochsenfeld(2008)}]{Beer2008}
\bibinfo{author}{M.~Beer}, \bibinfo{author}{C.~Ochsenfeld},
  \bibinfo{journal}{The Journal of Chemical Physics} \bibinfo{volume}{128}
  (\bibinfo{year}{2008}) \bibinfo{pages}{221102}. \URLprefix
  \url{http://aip.scitation.org/doi/10.1063/1.2940731}.
  \DOIprefix\doi{10.1063/1.2940731}.
%Type = Article
\bibitem[{Kussmann and Ochsenfeld(2007)}]{Kussmann2007}
\bibinfo{author}{J.~Kussmann}, \bibinfo{author}{C.~Ochsenfeld},
  \bibinfo{journal}{The Journal of Chemical Physics} \bibinfo{volume}{127}
  (\bibinfo{year}{2007}) \bibinfo{pages}{204103}. \URLprefix
  \url{http://aip.scitation.org/doi/10.1063/1.2794033}.
  \DOIprefix\doi{10.1063/1.2794033}.

\end{thebibliography}

\section{Acknowledgments}
This work was supported by CARCH4205, and by the Strategic Priority Research Program of Chinese Academy of Sciences~(Grant No. XDC01040100). 
H.S. acknowledges Victor Wen zhe Yu for inspiring discussions. H.S. thanks the Tianhe-2 Supercomputer Center for computational resources.
%We further acknowledge Volker Blum for his continued support during this project.

%Supported by the Strategic Priority Research Program of Chinese Academy of Sciences,Grant No. XDC01000000.

\appendix
\section{Appendix: Validation of the polarizability tensor for molecules}
\label{appendix:mol}
We use the linear scaling TC2-CPSCF method described in Sec.~{\ref{subsec:ON} to calculate the  polarizabilities of 32 selected molecules, the results are compared with the normal $O(N^3)$ method 
described in Sec.~{\ref{subsec:ON3} to serve as the benchmark to make the validation.  All calculations were performed at the LDA level of theory and using ``tier 2'' basis sets with the ``really tight'' defaults for the integration grids. The NTPoly-filter is set to $10^{-8}$ and NTPoly-tolerance is set to $10^{-4}$. The employed  equilibrium geometries were determined by relaxation that all absolute forces are smaller than $10^{-4}$~eV/$\mbox{\AA}$.  As summarized in Table~\ref{tab:dimers}, the
 mean absolute error (MAE) and the mean absolute percentage error (MAPE) is 0.023~Bohr$^{3}$ and 0.078\%. The largest  absolute error~(0.68~Bohr$^{3}$) observed in the LiH molecule, and this is  
because the density matrix purification convergence is not tight enough, if we change the NTPoly-tolerance to $10^{-6}$, then the largest absolute error in LiH is reduced to 0.0001~Bohr$^{3}$.
     % 29.86768  29.86768  30.63383
For the other larger molecules as shown in Table~\ref{tab:trimers_and_mol}, the mean absolute error (MAE) and the mean absolute percentage error (MAPE) is 0.0036~Bohr$^{3}$ and 0.015\%, which show an excellent agreement between our linear scaling TC2-DFPT implementation and the benchmark $O(N^3)$ DFPT results. 

\begin{table}
\scalebox{0.8}
{
\begin{tabular}{c c | c c c c }
\hline \hline
   &  &   TC2-CPSCF  &   benchmark &      ab-err~(Bohr$^{3}$) &  rel-err(\%) \\
\hline
Cl$_2$ &   $\alpha_{xx}$   & 24.12266 & 24.12271 & 0.00005 & 0.00021 \\
&   $\alpha_{yy}$   & 24.12266 & 24.12271 & 0.00005 & 0.00021 \\
&   $\alpha_{zz}$   & 41.31822 & 41.30860 & 0.00962 & 0.02328 \\
ClF &   $\alpha_{xx}$   & 15.92653 & 15.92653 & 0.00000 & 0.00000 \\
&   $\alpha_{yy}$   & 15.92653 & 15.92653 & 0.00000 & 0.00000 \\
&   $\alpha_{zz}$   & 22.29437 & 22.29239 & 0.00198 & 0.00888 \\
CO &   $\alpha_{xx}$   & 11.65557 & 11.65559 & 0.00002 & 0.00017 \\
&   $\alpha_{yy}$   & 11.65557 & 11.65559 & 0.00002 & 0.00017 \\
&   $\alpha_{zz}$   & 15.49260 & 15.49262 & 0.00002 & 0.00013 \\
CS &   $\alpha_{xx}$   & 22.23854 & 22.23855 & 0.00001 & 0.00004 \\
&   $\alpha_{yy}$   & 22.23854 & 22.23855 & 0.00001 & 0.00004 \\
&   $\alpha_{zz}$   & 37.65264 & 37.65202 & 0.00062 & 0.00165 \\
F$_2$ &   $\alpha_{xx}$   & 6.16984 & 6.16981 & 0.00003 & 0.00049 \\
&   $\alpha_{yy}$   & 6.16984 & 6.16981 & 0.00003 & 0.00049 \\
&   $\alpha_{zz}$   & 11.68810 & 11.68369 & 0.00441 & 0.03773 \\
H$_2$ &   $\alpha_{xx}$   & 3.89990 & 3.90139 & 0.00149 & 0.03821 \\
&   $\alpha_{yy}$   & 3.89990 & 3.90139 & 0.00149 & 0.03821 \\
&   $\alpha_{zz}$   & 7.54590 & 7.53197 & 0.01393 & 0.18460 \\
HCl &   $\alpha_{xx}$   & 16.81678 & 16.81511 & 0.00167 & 0.00993 \\
&   $\alpha_{yy}$   & 16.81678 & 16.81511 & 0.00167 & 0.00993 \\
&   $\alpha_{zz}$   & 18.88517 & 18.86799 & 0.01718 & 0.09097 \\
HF &   $\alpha_{xx}$   & 4.96469 & 4.96371 & 0.00098 & 0.01974 \\
&   $\alpha_{yy}$   & 4.96469 & 4.96371 & 0.00098 & 0.01974 \\
&   $\alpha_{zz}$   & 6.41086 & 6.40950 & 0.00136 & 0.02121 \\
Li$_2$ &   $\alpha_{xx}$   & 120.63028 & 120.63041 & 0.00013 & 0.00011 \\
&   $\alpha_{yy}$   & 120.63028 & 120.63041 & 0.00013 & 0.00011 \\
&   $\alpha_{zz}$   & 231.99591 & 231.98644 & 0.00947 & 0.00408 \\
LiF &   $\alpha_{xx}$   & 11.16256 & 11.16231 & 0.00025 & 0.00224 \\
&   $\alpha_{yy}$   & 11.16256 & 11.16231 & 0.00025 & 0.00224 \\
&   $\alpha_{zz}$   & 11.07222 & 11.06375 & 0.00847 & 0.07650 \\
LiH &   $\alpha_{xx}$   & 29.88636 & 29.86767 & 0.01869 & 0.06254 \\
&   $\alpha_{yy}$   & 29.88636 & 29.86767 & 0.01869 & 0.06254 \\
&   $\alpha_{zz}$   & 31.31993 & 30.63369 & 0.68624 & 2.19106 \\
N2 &   $\alpha_{xx}$   & 9.92305 & 9.92305 & 0.00000 & 0.00000 \\
&   $\alpha_{yy}$   & 9.92305 & 9.92305 & 0.00000 & 0.00000 \\
&   $\alpha_{zz}$   & 15.03340 & 15.03338 & 0.00002 & 0.00013 \\
Na$_2$ &   $\alpha_{xx}$   & 121.13132 & 121.13152 & 0.00020 & 0.00017 \\
&   $\alpha_{yy}$   & 121.13132 & 121.13152 & 0.00020 & 0.00017 \\
&   $\alpha_{zz}$   & 283.94622 & 283.91531 & 0.03091 & 0.01089 \\
NaCl &   $\alpha_{xx}$   & 28.19794 & 28.15581 & 0.04213 & 0.14941 \\
&   $\alpha_{yy}$   & 28.19794 & 28.15581 & 0.04213 & 0.14941 \\
&   $\alpha_{zz}$   & 40.78853 & 40.55770 & 0.23083 & 0.56592 \\
P$_2$ &   $\alpha_{xx}$   & 34.72382 & 34.72382 & 0.00000 & 0.00000 \\
&   $\alpha_{yy}$   & 34.72382 & 34.72382 & 0.00000 & 0.00000 \\
&   $\alpha_{zz}$   & 67.28002 & 67.27964 & 0.00038 & 0.00056 \\
SiO &   $\alpha_{xx}$   & 24.57026 & 24.57026 & 0.00000 & 0.00000 \\
&   $\alpha_{yy}$   & 24.57026 & 24.57026 & 0.00000 & 0.00000 \\
&   $\alpha_{zz}$   & 34.02059 & 34.02057 & 0.00002 & 0.00006 \\
\hline
MAE &  &        &          & 0.023  &    \\
MAPE & &        &          &       & 0.078\% \\
\hline \hline
\end{tabular}}
\caption{Polarizability tensor elements $\alpha$ for 16 dimers, as computed with the presented TC2-CPSCF implementation at the LDA level of theory. Additionally, mean absolute errors (MAE) and mean absolute percentage errors (MAPE) with respect to benchmark DFPT calculations are given. }
\label{tab:dimers}
\end{table}

\begin{table}
\scalebox{0.8}
{
\begin{tabular}{c c | c c c c }
\hline \hline
     &  &   TC2   &   benchmark &      ab-err~(Bohr$^{3}$) &  rel-err(\%) \\
\hline 
CO$_2$ &   $\alpha_{xx}$   & 12.04009 & 12.04059 & 0.00050 & 0.00415 \\
&   $\alpha_{yy}$   & 12.04009 & 12.04059 & 0.00050 & 0.00415 \\
&   $\alpha_{zz}$   & 26.55813 & 26.55857 & 0.00044 & 0.00166 \\
H$_2$O &   $\alpha_{xx}$   & 8.57558 & 8.57592 & 0.00034 & 0.00396 \\
&   $\alpha_{yy}$   & 9.79467 & 9.79485 & 0.00018 & 0.00184 \\
&   $\alpha_{zz}$   & 9.19050 & 9.19066 & 0.00016 & 0.00174 \\
HCN &   $\alpha_{xx}$   & 13.10165 & 13.10070 & 0.00095 & 0.00725 \\
&   $\alpha_{yy}$   & 13.10165 & 13.10070 & 0.00095 & 0.00725 \\
&   $\alpha_{zz}$   & 23.10572 & 23.10246 & 0.00326 & 0.01411 \\
SH$_2$ &   $\alpha_{xx}$   & 23.16771 & 23.16910 & 0.00139 & 0.00600 \\
&   $\alpha_{yy}$   & 24.10855 & 24.10932 & 0.00077 & 0.00319 \\
&   $\alpha_{zz}$   & 24.05320 & 24.05222 & 0.00098 & 0.00407 \\
SO$_2$ &   $\alpha_{xx}$   & 18.86849 & 18.86846 & 0.00003 & 0.00016 \\
&   $\alpha_{yy}$   & 33.63386 & 33.63384 & 0.00002 & 0.00006 \\
&   $\alpha_{zz}$   & 22.71019 & 22.70982 & 0.00037 & 0.00163 \\
C$_2$H$_2$ &   $\alpha_{xx}$   & 16.32317 & 16.32319 & 0.00002 & 0.00012 \\
&   $\alpha_{yy}$   & 16.32317 & 16.32319 & 0.00002 & 0.00012 \\
&   $\alpha_{zz}$   & 31.80234 & 31.80238 & 0.00004 & 0.00013 \\
C$_2$H$_4$ &   $\alpha_{xx}$   & 20.20784 & 20.20808 & 0.00024 & 0.00119 \\
&   $\alpha_{yy}$   & 24.66562 & 24.66577 & 0.00015 & 0.00061 \\
&   $\alpha_{zz}$   & 35.70510 & 35.70541 & 0.00031 & 0.00087 \\
CH$_3$Cl &   $\alpha_{xx}$   & 26.34065 & 26.32743 & 0.01322 & 0.05019 \\
&   $\alpha_{yy}$   & 26.34066 & 26.32743 & 0.01323 & 0.05023 \\
&   $\alpha_{zz}$   & 36.02488 & 35.99897 & 0.02591 & 0.07192 \\
CH$_4$ &   $\alpha_{xx}$   & 16.97370 & 16.97375 & 0.00005 & 0.00029 \\
&   $\alpha_{yy}$   & 16.97370 & 16.97375 & 0.00005 & 0.00029 \\
&   $\alpha_{zz}$   & 16.97370 & 16.97375 & 0.00005 & 0.00029 \\
H$_2$CO &   $\alpha_{xx}$   & 11.99309 & 11.99307 & 0.00002 & 0.00017 \\
&   $\alpha_{yy}$   & 18.33276 & 18.33274 & 0.00002 & 0.00011 \\
&   $\alpha_{zz}$   & 23.03242 & 23.03234 & 0.00008 & 0.00035 \\
H$_2$O$_2$&   $\alpha_{xx}$   & 13.60503 & 13.59588 & 0.00915 & 0.06725 \\
&   $\alpha_{yy}$   & 17.61033 & 17.59463 & 0.01570 & 0.08915 \\
&   $\alpha_{zz}$   & 12.36691 & 12.36236 & 0.00455 & 0.03679 \\
N$_2$H$_4$ &   $\alpha_{xx}$   & 20.99202 & 20.99657 & 0.00455 & 0.02167 \\
&   $\alpha_{yy}$   & 25.92763 & 25.88185 & 0.04578 & 0.17657 \\
&   $\alpha_{zz}$   & 21.20460 & 21.20944 & 0.00484 & 0.02283 \\
NH$_3$ &   $\alpha_{xx}$   & 13.34055 & 13.33894 & 0.00161 & 0.01207 \\
&   $\alpha_{yy}$   & 13.34052 & 13.33895 & 0.00157 & 0.01177 \\
&   $\alpha_{zz}$   & 14.60966 & 14.60798 & 0.00168 & 0.01150 \\
PH$_3$ &   $\alpha_{xx}$   & 30.00257 & 30.00255 & 0.00002 & 0.00007 \\
&   $\alpha_{yy}$   & 30.00259 & 30.00257 & 0.00002 & 0.00007 \\
&   $\alpha_{zz}$   & 31.11588 & 31.11584 & 0.00004 & 0.00013 \\
Si$_2$H$_6$ &   $\alpha_{xx}$   & 57.44315 & 57.44406 & 0.00091 & 0.00158 \\
&   $\alpha_{yy}$   & 57.44294 & 57.44401 & 0.00107 & 0.00186 \\
&   $\alpha_{zz}$   & 77.03289 & 77.03454 & 0.00165 & 0.00214 \\
SiH$_4$ &   $\alpha_{xx}$   & 31.97412 & 31.96648 & 0.00764 & 0.02389 \\
&   $\alpha_{yy}$   & 31.97413 & 31.96648 & 0.00765 & 0.02393 \\
&   $\alpha_{zz}$   & 31.97412 & 31.96648 & 0.00764 & 0.02389 \\
\hline
MAE &  &        &          &  0.0036 &    \\
MAPE & &        &          &       & 0.015\% \\
\hline \hline
\end{tabular}}
\caption{Polarizability tensor elements $\alpha$ for 16 molecules, as computed with the presented TC2-CPSCF implementation at the LDA level of theory. Additionally, mean absolute errors (MAE) and mean absolute percentage errors (MAPE) with respect to benchmark DFPT calculations are given. }
\label{tab:trimers_and_mol}
\end{table}

\section{Appendix: The comparison for the total time} 
\label{appendix:total_time}
The comparison of the total time for the calculation of the polarizabilities between the $O(N)$ TC2-CPSCF method and the traditional $O(N^3)$ method is shown in Fig.~\ref{fig:a_compare}. 
 \begin{figure}
 \centering
 \includegraphics[width=0.9\columnwidth]{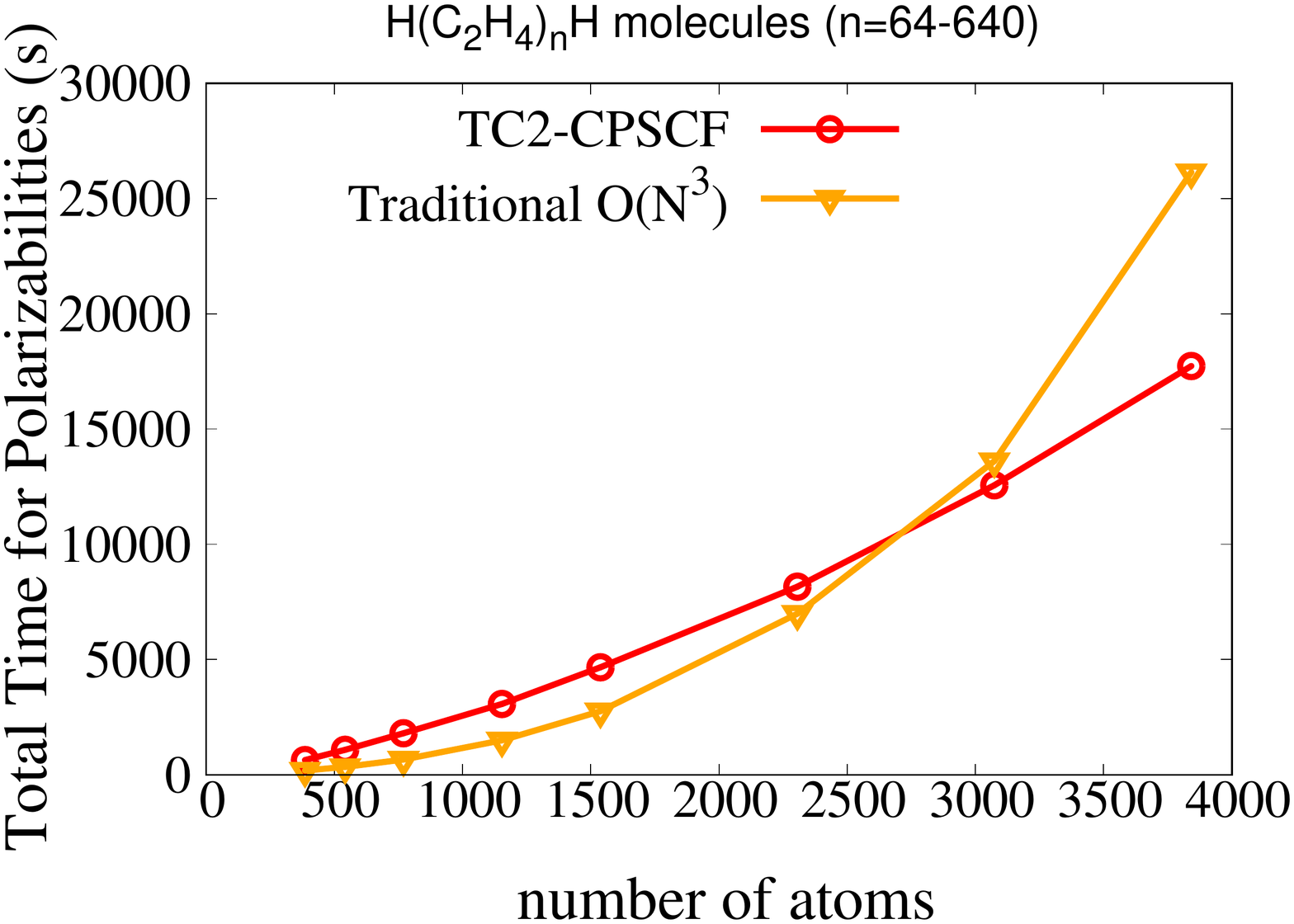}
 \caption{ The total time of the calculation of the polarizabilities for isolated H(C2H4)$_n$H molecules containing from 386 to 3842 atoms. Here we use NTPoly-filter~($10^{-6}$) and NTPoly-tolerance~($10^{-5}$) as the numerical thresholds. All calculations are performed on 36 CPU cores. }
 \label{fig:a_compare}
 \end{figure}

\end{document}